\documentclass[12pt,a4paper]{article}
\usepackage{subfig}
\usepackage{graphicx}
\usepackage{epsfig}
\usepackage{amssymb}

\begin{document}

\title{On the multifractal effects generated by monofractal signals}

\author{Dariusz Grech\footnote{dgrech@ift.uni.wroc.pl} and Grzegorz Pamu{\l}a\footnote{gpamula@ift.uni.wroc.pl} \\
Institute of Theoretical Physics\\
University of Wroc{\l}aw, PL-50-204 Wroc{\l}aw, Poland}
\date{}

\maketitle

\begin{abstract}
We study quantitatively the level of false multifractal signal one may encounter while analyzing multifractal phenomena in time series within multifractal detrended fluctuation analysis (MF-DFA). The investigated effect appears as a result of finite length of used data series and is additionally amplified by the long-term memory the data eventually may contain. We provide the detailed quantitative description of such apparent multifractal background signal as a threshold in spread  of generalized Hurst exponent values $\Delta h$ or a threshold in the width of multifractal spectrum  $\Delta \alpha$ below which multifractal properties of the system are only apparent, i.e. do not exist, despite $\Delta\alpha\neq0$ or $\Delta h\neq 0$. We find this effect quite important for shorter or persistent series and we argue it is linear with respect to autocorrelation exponent $\gamma$. Its strength decays according to power law with respect to the length of time series. The influence of basic linear and nonlinear transformations applied to initial data in finite time series with various levels of long memory is also investigated. This  provides additional set of semi-analytical results. The obtained formulas are significant in any interdisciplinary  application of multifractality, including physics, financial data analysis or physiology, because they allow to separate the 'true' multifractal phenomena from the apparent (artificial) multifractal effects. They should be a helpful tool of the first choice to decide whether we do in particular case with the signal with real multiscaling properties or not.

\end{abstract}

$$
$$

\vspace{1cm}

\section{Introduction}

The presence of long range memory in complex systems has been studied in the literature in a wide variety of fields with a hope of better understanding the dynamics of these systems and to make their evolution more predictive.
In the simplest case of stationary and monofractal time series the memory effects are defined with the use of two point autocorrelation function $C_s\sim\langle\Delta x_i \Delta x_{i+s}\rangle $, where $x_i$, $(i=1,...,L)$ are data in series, $L$ is its length,  $\langle \rangle $ is an average taken over all data in the given signal and $\Delta x_i = x_{i+1}-x_i$ are series increments.
It is well known that $C_s$ reveals in this case the power law form in an asymptotic limit of large time lags $s$ \cite{gamma_1,gamma_2}

\begin{equation}
C_s\simeq (2-\gamma)(1-\gamma)s^{-\gamma},
\end{equation}
where the scaling exponent $\gamma$ $(0\leq\gamma\leq1)$ is related with the level of memory in signal and describes at boundaries fully autocorrelated (persistent) signal for $\gamma =0$ or completely uncorrelated increments for $\gamma=1$.

A direct calculation of correlation function and $\gamma$ exponent gets more difficult in real signals when a noise or nonstationarities are present in data. Therefore, to avoid this problem, an alternative approach is often made where the fluctuations in the cumulated data, i.e. fluctuations in the series $x_t=\sum_{i=1}^t \Delta x_i$ $(t=1,...,L)$, are considered. One measures the scaling Hurst exponent $H$ \cite{Hurst1,Hurst2} of $x_t$ related to $\gamma$ according to \cite{rel_gamma_H}
\begin{equation}
H=1-\frac{\gamma}{2}
\end{equation}
This way, any measurement of $H$ ($0.5 \leq H \leq 1$) is mapped to the corresponding $\gamma$ values.

The most prominent example of such indirect analysis of memory effects is the detrended fluctuation analysis (DFA) \cite{DFA_2,DFA_1,DFA_3}. The scaling law from Eq.(1) is replaced within DFA by the power law
\begin{equation}
F(\tau)\simeq \tau^H
\end{equation}
where $F(\tau)$ is the averaged fluctuation of the signal around its local trend in  time windows of fixed length $\tau$.  One has in detailed quantitative description:
\begin{equation}
F(\tau)=\frac{1}{2N}\sum^{2N}_{k=1} \hat{F}^2(\tau,k)
\end{equation}
where
\begin{equation}
\hat{F}^2(\tau,k)=\frac{1}{\tau}\sum^{\tau}_{j=1}\left\{x_{(k-1)\tau+j}-P_k(j)\right\}^2
\end{equation}
Here $N=[L/{\tau}]$ stands for the number of non-overlapping boxes for which the detrending procedure is performed and $P_k$ is the polynomial trend fitted to initial data in $k$-th time window and then subtracted from these data.

An application of DFA is limited to systems whose memory properties are well described by a single scaling exponent. Such an approach is insufficient in so called multifractal systems \cite{mf-281-physrep}--\cite{Kantelhardt_encyklopedia}, where small and large fluctuations follow different scaling laws (see, e.g. \cite{rel_gamma_H,5}--\cite{MF-DFA}). One has to employ more general method of description in the latter case and a full understanding of issues connected with memory properties in multifractal data is still under debate. The multifractal properties of time series  are now
extensively studied because of the omnipresence of multifractals in various phenomena in nature. Its presence is confirmed in empirical data from turbulence \cite{mf-turb-1,mf-turb-2}, astronomy \cite{mf-astr}, climate phenomena \cite{mf-clima}, physiology \cite{mf-physiol}, text structure \cite{mf-text}, physics \cite{mf-phys-1,mf-phys-2} up to finances \cite{mf-finance-1}--\cite{mf-finance-5}, not covering many other publications on these phenomena. It is worth to stress that the answer to questions regarding accuracy, applicability and reliability of multifractal measurements is crucial for the development of efficient simulation or forecasting methods \cite{Kantelhardt_encyklopedia}.

One of the most frequently used techniques to quantify the multifractal properties in data series is multifractal detrended fluctuation analysis (MF-DFA) \cite{MF-DFA}.
The MF-DFA method enables to eliminate polynomial trends in data what makes it superior to other multifractal oriented techniques like  the structure
function analysis \cite{SFA} or the higher-order autocorrelation
functions \cite{HOAF}. The popular method to study multifractal properties is also the wavelet transform modulus maxima (WTMM) approach \cite{WTMM}. WTMM eliminates polynomial trends similarly to MF-DFA but is considerably more difficult to implement. Moreover, it may give biased outcomes and yields spurious multifractality more often \cite{WTMM-MFDFA}.

The MF-DFA has been applied so far in diversified scientific problems, just to mention:
seismology \cite{MF-seismology_1,MF-seismology_2}, cosmology \cite{MF-cosmology}, biology \cite{MF-biology_1, MF-biology_2},
meteorology \cite{MF-meteo}, medicine \cite{MF-medicine},
music \cite{MF-music_1,MF-music_2}, geophysics \cite{MF-geophys}, and finances \cite{MF-finance_7}--\cite{czarnecki} far from being exhaustive here.
Let us remind that the main ingredient of MF-DFA is the $q$-weighted fluctuation $F_q(\tau)$ of the time series signal around its local trend  in a time window of size $\tau$. More precisely, it is defined as an $q$-th moment ($q \in \mathbf{R}$) of $F(\tau)$ already used in DFA, i.e.
\begin{equation}
F_q(\tau)=\left\{\frac{1}{2N}\sum^{2N}_{k=1} [\hat{F}^2(\tau,k)]^{q/2}\right\}^{1/q}
\end{equation}
for $q\neq 0$, and
\begin{equation}
F_0(\tau)=\exp\left\{\frac{1}{4N}\sum^{2N}_{k=1} \ln[\hat{F}^2(\tau,k)]\right\}
\end{equation}
for $q=0$.

The power law relation
\begin{equation}
F_q(\tau) \sim \tau^{h(q)}
\end{equation}
is expected in MF-DFA which forms a basis for the so called generalized Hurst exponent $h(q)$ \cite{MF-DFA}. Eq.(8) coincides with Eq.(3) for $q=2$, hence $h(2)=H$ returns the main Hurst exponent value.

The multifractal properties may  alternatively be presented as the multifractal spectrum $(\alpha, f(\alpha))$ \cite{mf-281-physrep}, known also as H{\"o}lder description.
Both descriptions are linked together via relations \cite{legendre1,legendre2}
\begin{equation}
\alpha(q)=h(q)+qh'(q),\qquad f(\alpha) = q(\alpha-h(q))+1
\end{equation}

One usually calculates the strength of multifractality
by the spread of $h(q)$ profile, i.e. $\Delta h= h(-q) - h(+q)$ , ($q>0$) or by the width of multifractal spectrum $f(\alpha)$, i.e.
$\Delta \alpha=\alpha(-q) - \alpha(+q)$. Both quantities coincide in the limit $q\rightarrow\infty$. In the case of stationary signal the $h(q)$ profile has a monotonic behavior \cite{MF-DFA} and
the spectrum $f(\alpha)$ has a form of inverted parabola. Its spread corresponds to the $\Delta \alpha$ value.
Through this article two above descriptions will be used.

It is worth to stress that multifractality is an interesting and perspective phenomenon for further investigation and
applications. On the other hand, it is very subtle and generally difficult task to distinguish in many real situations between
'true' multifractal properties of a system and the 'apparent' multifractal phenomena resulting directly from the records based on MF-DFA. Recently, some new features connected
with the so called spurious and corrupted multifractality were shown \cite{spur-corr-MF}--\cite{spur-corr-MF_3}. It turns out that the presence of
additive white or color noise, short-term memory or periodicities in multifractal signals may
significantly change its observed multifractal properties for all data lengths. Such contamination by additions of
disturbances are typically found in many observational records in various complex systems: in physics, finance,
physiology, geology, climate dynamics, network traffic, etc.
All real series contain also finite sets of data. 
Due to smaller statistics in short signals, the accidental fluctuations which are not related with multifractal properties contribute also to fluctuation function $F_q$ and may even dominate over fluctuations related with multifractality.
On the other hand, the influence of large fluctuations in short
time series can be suppressed since MF-DFA algorithm may not distinguish them from longer
trends which are eliminated.
 In both cases the observed $\Delta h$ spread is increased comparing
with the corresponding value obtained for much longer signals.
This raises a question about confidence for multifractal findings which are naturally affected by finite size effects in autocorrelated or
uncorrelated data even if they are not contaminated by the additional effects mentioned before.

The nonlinear transformations of primary data also significantly affect the registered multifractal properties usually magnifying them. Therefore, a
link between 'true' multifractal picture of the complex system and the observed quantitative properties which may suggest the presence of multifractality, even if it is absent in the system,  is still an intriguing and
not fully understood issue. It will obviously depend on the technique used to quantify multifractal phenomena.
In this paper we use MF-DFA as the commonly accepted technique to find multifractal properties of time series.

Our goal is a deeper investigation of finite size effects (FSEs) influencing within MF-DFA method the observed multifractal features
of time series.
The finite size effects in multifractality were mentioned already in Ref. \cite{lux-ausloos}. Since then several authors approached this problem (see e.g. Refs. \cite{schu-kantelhardt,H8_27,zhou}), however general search of this phenomenon has not been yet investigated.
The particular emphasis will be given to amplification of an apparent observed multifractal effects caused by the presence of long memory in signals and by basic transformations done on initial monofractal data.
These effects should be distinguished from
the spurious or corrupted multifractality \cite{spur-corr-MF,spur-corr-MF_3}, where the given data series contaminated by various
effects like noise,
short-term memory, periodicities in signal, etc., change the shape of multifractal profile $h(q)$ or $f(\alpha)$ spectrum. 
The latter effects will affect multifractality for all lengths of data $L$,
while FSE should disappear if $L\rightarrow\infty$.
Note, that one always deals in practice with finite set
of data, so one should exactly know the level of confidence for results obtained within MF-DFA if they indicate a multifractality in the system, i.e. $\Delta h\neq 0$ or $\Delta \alpha\neq 0$.
The preliminary quantitative analysis of this problem had already been made in our conference paper \cite{FENS}.

Multifractal finite time series will contain the multifractal FSE bias (i.e., the threshold spread of
generalized Hurst exponent values not related with multifractal properties of these series) at least like a lot less complicated
monofractal series of similar length and persistency level. We will deal with simpler, involving fewer parameters and more
clear for synthetic generation monofractal data that will serve as a tool
to find the minimal multifractal FSE bias threshold in all kind of signal.
We start with description of FSE in monofractal series with the analysis of generalized Hurst exponent $h(q)$.
Then this description is extended to an alternative approach of multifractal spectra in terms of $\Delta \alpha$ in Section 3. Both sections provide detailed analysis of threshold values $\Delta h$ and $\Delta \alpha$ below which the multifractal properties of complex system are only apparent. The corresponding FSE multifractal phenomena for linearly transformed data are discussed in Section 4. An interesting and important from practical point of view phenomenon of multifractal bias generated solely by nonlinear transformations in monofractal data is described in Section 5. One has to know the strength of of such bias to get the evidence of multifractal properties in primary not transformed data (and in complex system in general) if only volatile series of data from such systems are available. Section 6. summarizes semi-analytical results obtained in this article and contains final remarks and conclusions.

\section{Properties of generalized Hurst exponents in finite\\ monofractal signals}

We start to describe quantitatively, in terms of generalized Hurst exponents \cite{MF-DFA}, the expected level of multifractal
background existing due to the finite-size effects for uncorrelated data as well as for time series
with long memory induced by the explicit form of autocorrelations. Our approach is based
on the Fourier filtering method (FFM) \cite{FFM} that directly shapes the memory level in artificial data by the respective choice of autocorrelation exponent $\gamma$ (see Eq.(1)). In order to check the accuracy of FFM for generation of time series with a given persistency level, we examined this procedure towards replication of the pre-assumed autocorrelation properties in artificially constructed series. This efficiency is demonstrated in Fig.1. 
The scaling resulting from Eq. (8) was found also very satisfactory (see Fig.2)

The performance of MF-DFA technique strictly depends on the power law scaling  between  $q-\mathrm{th}$ order fluctuations $F_q(\tau)$ and the box size $\tau$ (see Eq.(8)). An exact extraction of the generalized Hurst exponent $h(q)$ requires well determined scaling range for the linear fit $\log F_q(\tau)$ against $\log \tau$, induced by  the power law in Eq.(8). The expected power law dependence for various lengths of the signal $L$ and for different values of deformation parameter $q$ is shown in Fig.2. The scaling range from $\tau=20$ till $ \tau=L/4$, confirmed by plots in Fig.2, and the second degree polynomial trends $P_k$ were chosen to be used further on.

We considered the ensembles of numerically generated monofractal time series of length $L=2^n$, $(n=9, 10,\ldots, 20)$ with the pre-assumed autocorrelation exponent value $\gamma = 0.1, 0.2,\ldots, 0.9, 1.0$, each containing $10^2$ independent realizations.
Thus, the spread of $\gamma$ exponents covered uniformly the range $1/2\leq H \leq 1$. Every obtained quantity has been averaged over such statistical ensemble. The width $\Delta h$ of the generalized Hurst
exponents profile $h(q)$ was limited to the edge values $q=\mp 15$. Notice that the range $-15\leq q\leq +15$ is large enough for checking multiscaling properties of data from large to considerably very small fluctuations
(see Fig.3).

The key numerical data  are presented in series of plots in Figs.4-7, where $h^{\pm}\equiv h(\pm 15)$ and
$\Delta h\equiv h^{-}-h^{+}$, ($h^{-}>h^{+}$).
First, the uncorrelated artificial data were studied. They have been generated from Gaussian distribution ($\gamma=1$),
and also independently obtained as the shuffle of autocorrelated signals. Fig.4a shows a dependence of
the edge values $h^{\pm}$ of the generalized Hurst exponents $h(q)$ for two different lengths of time
series: $L = 2^{12}$ and $L=2^{20}$ chosen to investigate short and long data sets respectively. These plots not only confirm that shuffling procedure leads
to the same results as Gaussian distribution does (what should be expected anyway), but also reveal the dependence
of  $h^{\pm}$ values on the data length. The corresponding plots for the intermediate lengths are not shown for they look qualitatively
the same, i.e. $h^{\pm}$  does not change with $\gamma$ for shuffled data but it does depend on $L$.

This way, the spread $\Delta h$  of generalized Hurst exponent versus length of data can be investigated
for signals with no memory (see Fig.5a). The data drawn in log-log scale convince about the power law dependence between $\Delta h_1\equiv\Delta h(\gamma=1)$ and the length of data series $L$

\begin{equation}
\Delta h_1(L) = C_1L^{-\eta_1}
\end{equation}
with some real constants $C_1$ and $\eta_1$.

Next, keeping $L$
fixed, we turned to investigate  the edge values for $h(\pm15)$ versus the autocorrelation exponent value $\gamma$
for monofractal signals  with long memory. The examples of this dependence are shown  in Fig.6a for chosen lengths $L = 2^{12}$ and $L = 2^{20}$. The cases of
remaining lengths (not shown due to lack of space) have been checked by us as well and look similarly. All of them indicate an existence of  the excellent linear relationship between $h(\pm15)$ and
$\gamma$ in the whole range of $\gamma$ values.
Therefore, we get
\begin{equation}
\Delta h(\gamma,L) = A(L)\gamma + B(L)
\end{equation}
where the coefficients $A(L)$ and $B(L)$ are functions of $L$ only.

The boundary conditions, i.e. the form of  $\Delta h_1(L)$ and $\Delta h_0(L)\equiv \Delta h(\gamma=0, L)$
allow to specify the final shape of this relation.
The profile  $\Delta h_0(L)$ cannot be found directly, because one is stuck with singularity in FFM method for $\gamma=0$.
Therefore we made the extrapolation of the fitting lines $h(\pm 15)$ versus $\gamma$ up to the point $\gamma\rightarrow 0$
for all lengths to get  the collection of $\Delta h_0(L)$ values. It is shown in Fig.6a.
The plot in log-log scale against the length of time series in Fig.7a confirms that for fully persistent time series ($\gamma\rightarrow 0, H\rightarrow 1$) $\Delta h_0(L)$
gets also the power law form:
\begin{equation}
\Delta h_0(L)=C_0L^{-\eta_0}
\end{equation}
with some constants $C_0$ and $\eta_0$ to be determined from the fit.

Inserting Eq.(10) and Eq.(12) into Eq.(11), we arrive with the final formula for $\Delta h(\gamma, L)$
\begin{equation}
\Delta h(\gamma,L)=C_1 L^{-\eta_1}\gamma + C_0 L^{-\eta_0} (1-\gamma)
\end{equation}

In practical applications, the shape of the $95\%$ confidence level profile is important for the multifractal bias
caused by FSE and amplified by the presence of long memory in data. It shall be given by the same formula but with
different coefficients calculated on the basis of $1\sigma$ uncertainties
$\sigma_{C_p}$,  $\sigma_{\eta_p}$ associated with the fit of $C_p$ and $\eta_p$ parameters, where $p=0,1$
corresponds to boundary conditions at $\gamma=0$ and $\gamma=1$ respectively.
One has  to include also
the standard deviation $S_p$ resulting from the series statistics. The formulas given in Eq.(10) and Eq.(12) have to be
replaced then by

\begin{equation}
\Delta h_p^{95\%}(L) = C_p\exp(f(\sigma_{C_p}+S_p))L^{-{\eta}_p+f\sigma_{\eta_p}}
\end{equation}
with $f=1.65$ taken for the particular $95\%$ confidence level.

Introducing the notation
\begin{equation}
C_p^{95\%}=C_p\exp(f(\sigma_{C_p}+S_p))
\end{equation}
and
\begin{equation}
\eta_p^{95\%}=\eta_p-f\sigma_{\eta_p}
\end{equation}
one gets the $95\%$ confidence level profile for the multifractal FSE bias as

\begin{equation}
\Delta h^{95\%}(\gamma,L)=C^{95\%}_1 L^{-\eta^{95\%}_1}\gamma + C^{95\%}_0 L^{-\eta^{95\%}_0} (1-\gamma)
\end{equation}

The results of the best fit, done on the statistics of $10^2$ monofractal time series for all quoted coefficients with uncertainties entering Eq.(15) and Eq.(16) are indicated in Table 1.

These findings may also be summarized as in Fig.8. They show FSE multifractal thresholds calculated at $95\%$ confidence level for $\Delta h$ spread in monofractal data series as a function of persistency level indicated by $\gamma$ value and for variety of data lengths $L$. The dotted horizontal lines present thresholds connected entirely with FSE, not affected by eventual autocorrelations in data. The  continuous sloped straight lines describe FSE multifractal thresholds amplified by the presence of long-term memory in series and predicted by Eq.(17). The 'true' multifractality, related with the existence of variety of scaling exponents in infinite system, may occur for consecutive series of given length only when the spread in generalized Hurst exponents enters the area above this line (at $95\%$ confidence level). This result clarifies and generalized quantitatively the statement from Ref.\cite{schu-kantelhardt} on the existence of such threshold values.

\begin{table}[!h]
\centering
\begin{tabular}{||c|c|c|c||c|c|c|c||} \hline
$C_1$ & $\eta_1$ & $C_0$ & $\eta_0$ & $C^{95\%}_1$ & $\eta^{95\%}_1$ & $C^{95\%}_0$ & $\eta^{95\%}_0$\\ \hline
0.603 & 0.175 & 0.453 & 0.124 & 0.631 & 0.171 & 0.484 & 0.120\\ \hline
\end{tabular}
\caption{Collected results for coefficients  of the fit describing the multifractal FSE thresholds in persistent time series according to Eqs.(13) and (17). The ensemble of $10^2$ simulations of time series was considered.}
\label{tab1}
\end{table}

\section{Finite size effects in analysis of multifractal spectrum}
All findings in the previous section can be easily translated into
$f(\alpha)$ singularity spectrum properties.
The multifractal spectrum width $\Delta \alpha$ is considered in literature as another useful measure of multifractality. Analogically to the \(\Delta h(L,\gamma)\) analysis, one may ask for the dependence of multifractal spectrum width \(\Delta\alpha\) on the signal length $L$ and on its persistency level $\gamma$. These results are obtained with the use of Eq.(9) applied to previously discussed and calculated generalized Hurst exponent values. Thus, the results should lead to similar qualitative conclusions and quantitatively they might be also valuable from the practical point of view.

The examples of multifractal spectrum ($\alpha, f(\alpha)$) for finite monofractal signals are shown in Fig.3b. Three cases: for strongly autocorrelated, medium autocorrelated, and weakly autocorrelated signals are indicated there for two distinct lengths of data: $L=2^{12}$ and $L=2^{20}$.
As previously, the first step is to examine the $\Delta\alpha$ characteristics obtained for randomly shuffled signal ($\gamma=1$). Fig.4b shows the minimal $\alpha_{min}$ and maximal $\alpha_{max}$ value of $\alpha$ parameter revealing lack of its dependence on $\gamma$. This proves again that shuffling procedure was efficient enough. The dependence $\Delta\alpha_1\equiv\Delta\alpha(\gamma=1)  $  on $L$ obeys a power law relation

\begin{equation}
\Delta\alpha_1(L)=D_1L^{-\xi_1},
\end{equation}
shown in Fig.5b, with constants $D_1$ and $\xi_1$ to be fitted. The 95\% confidence level for this relation is given by equation analogical to Eq.(14).

The edge values for $\alpha_{min/max}$ regarded as a function of $\gamma$ are presented  for particular lengths $L=2^{12}$ and $L=2^{20}$ in Fig.6b to confront them with previous findings for $h(\pm 15)$ in Fig.6a. The linear dependence for all other lengths (not shown) was also observed. Once we repeat the same approach to $\Delta\alpha(L,\gamma)$ dependence as we did before in section 2 and take into account the second boundary condition $\Delta\alpha_0(L) \equiv \Delta\alpha(L,\gamma=0)$ (see Fig.7b), we arrive with the final formula describing the character of multifractal spectrum width, similar to the one in Eq.(13), i.e.

\begin{equation}
\Delta \alpha(\gamma,L)=D_1 L^{-\xi_1}\gamma + D_0 L^{-\xi_0} (1-\gamma).
\end{equation}
The extension of this formula indicating the $95\%$ confidence level for FSE multifractal threshold, reproducing the result of Eq.(17) will be read respectively
\begin{equation}
\Delta \alpha^{95\%}(\gamma,L)=D^{95\%}_1 L^{-\xi^{95\%}_1}\gamma + D^{95\%}_0 L^{-\xi^{95\%}_0} (1-\gamma)
\end{equation}
The values of fitted coefficients for FSE multifractal thresholds in H\H{o}lder description for persistent and uncorrelated data are gathered in Table 2. The corresponding thresholds are indicated in plots of Fig.8 (see the right axis).

\vspace{2em}
\begin{table}[!h]
\centering
\begin{tabular}{||c|c|c|c||c|c|c|c||}
\hline
$D_1$ & $\xi_1$ & $D_0$ & $\xi_0$ & $D^{95\%}_1$ & $\xi^{95\%}_1$ & $D^{95\%}_0$ & $\xi^{95\%}_0$\\ \hline
0.686 & 0.129 & 0.572 & 0.089 & 0.784 & 0.120 & 0.670 & 0.079\\ \hline
\end{tabular}
\caption{Results for coefficients  of the fit describing the multifractal FSE threshold in terms of multifractal spread $\Delta \alpha$ for persistent and uncorrelated time series (see Eqs.(19) and (20)). The ensemble of $10^2$ simulations of time series was considered.}
\label{tab2}
\end{table}
\vspace{2em}

\section{Multifractal finite size effects  for linearly transformed data}

In many records of realistic data, a direct analysis of primary time series $x_t$ within power laws following from Eq.(3) or Eq.(8) may be difficult. For example, it is a case when $x_t$ contain some larger fluctuations or nonstationarities. A discussion of cumulated or integrated (in continuous case) data, i.e. $X_T\equiv\sum_{t=1}^T x_t$, usually reduces the scale of problems we have to face with. On the other hand, there are realistic time series very slowly changing in time. It is more convenient in such situation to study its inner structure directly looking at series increments $\Delta x_t=x_t-x_{t-1}$ instead of $x_t$ alone. Other more complex nonlinear transformations of initial data are also realized in some applications (e.g. volatility in finance).
One may therefore ask how the above transformations affect the formation of multifractal  FSE  bias described in previous sections and in particular, the recorded multifractal properties of data series after nonlinear operations.
We discuss in this section the problem of integration and differentiation of data, both meant in a discrete manner. The case of basic nonlinear transformations is considered in the following section, while an extension to more complex nonlinear transformations is treated numerically in our separate publication \cite{APPA-2013}.

The analysis was done for the same set of input parameters shaping the persistency level of initial data before transformation, i.e. $\gamma=0.1,0.2,..,0.9,1.0$ and with an identical statistical ensemble of  $10^2$ simulated time series of fixed length ($L=2^{10},2^{12},...,2^{18},2^{19},2^{20}$).
To distinguish quantities obtained for transformed series from the primary quantities calculated for data before transformation let us introduce indices $ (c)$ and $(d)$ corresponding respectively to integration or differentiation of data in primary series.
The edge values of the generalized Hurst exponent  obtained for monofractal series after integration ($h^{c\pm}$) or differentiation ($h^{d\pm}$) are presented altogether with the result for original series repeated from Fig.6. They are drawn as a function of $\gamma$ exponent for three artificially generated series from FFM of length $L=2^{12}, 2^{16}$ and $2^{20}$ (see Fig.9). We see that the linear dependence on $\gamma$ from Eq.(13) is reproduced

\begin{equation}
\Delta h^{c(d)}(\gamma, L) = A^{c(d)}(L)\gamma + B^{c(d)}(L)
\end{equation}
with new coefficients $A^{c(d)}(L)$ and $B^{c(d)}$ for both transformations.

 It is also well noticed that the spread of generalized Hurst exponent is much wider for integrated series and remarkably more narrow for differentiated series in comparison with the original primary data. The boundary conditions for the formula in Eq.(21) can be found  in the same manner as before in section 3. For fully autocorrelated data, an extrapolation  to $\gamma \rightarrow 0$ was used in Fig.9. This way we are able to reveal $\Delta h_0^{c(d)}(L)$ and $\Delta h_1^{c(d)}(L)$ relations, shown in Figs.10,11, which have the same power law form as for not transformed data in Eqs.(10) and (12), i.e.

\begin{equation}
\Delta h^{c(d)}_0(L) = C^{c(d)}_0L^{-\eta^{c(d)}_0}
\end{equation}
and
\begin{equation}
\Delta h^{c(d)}_1(L) = C^{c(d)}_1L^{-\eta^{c(d)}_1}
\end{equation}
thus leading to relationships
\begin{equation}
A^{c(d)}(L) = C^{c(d)}_1L^{-\eta^{c(d)}_1} - C^{c(d)}_0L^{-\eta^{c(d)}_0}
\end{equation}

\begin{equation}
B^{c(d)}(L) =  C^{c(d)}_0L^{-\eta^{c(d)}_0}
\end{equation}

The fitting results for coefficients in the above relations are collected in Table 3 for integrated series and in Table 4 for differentiated series. The results for $95\%$ confidence level, calculated in an analogical way as in section 2, are also indicated. These values fully determine the FSE multifractal profile of transformed monofractal series which may be summarized after simple replacement of corresponding coefficients in Eqs.(13) and (17) by  $C_p^{c(d)}$ and $\eta_p^{c(d)}$.

\begin{table}[!h]
\centering
\begin{tabular}{||c|c|c|c||c|c|c|c||}
\hline
$C^{c}_1$ & $\eta^{c}_1$ & $C^{c}_0$ & $\eta^{c}_0$ & $C^{c (95\%)}_1$ & $\eta^{c (95\%)}_1$ & $C^{c (95\%)}_0$ & $\eta^{c (95\%)}_0$\\ \hline
0.580&0.101&0.701&0.108&0.813& 0.122& 0.736& 0.112\\
\hline
\end{tabular}
\caption{Results of the fit for coefficients in Eqs.(22) and (23) describing multifractal FSE bias for integrated data done on ensemble of $10^2$ independent realizations.}
\label{tab3}
\end{table}

\begin{table}[!h]
\centering
\begin{tabular}{||c|c|c|c||c|c|c|c||}
\hline
$C^{d}_1$ & $\eta^{d}_1$ & $C^{d}_0$ & $\eta^{d}_0$ & $C^{d (95\%)}_1$ & $\eta^{d (95\%)}_1$ & $C^{d (95\%)}_0$ & $\eta^{d (95\%)}_0$\\ \hline
133.565 &0.885&7.753 &0.539&134.267&0.952& 7.930&0.555\\
\hline
\end{tabular}
\caption{Results of the fit for coefficients in Eqs.(22) and (23) describing multifractal FSE bias for differentiated data done on ensemble of $10^2$ independent realizations.}
\label{tab4}
\end{table}

A translation of these findings into H\H{o}lder singularity spectrum properties is straightforward. Once repeating the consecutive steps we arrive with series of twin plots shown in right-hand side of Figs.9-11. They finally lead to the similar multifractal FSE bias formula in the H\H{o}lder language, linking the expected level of FSE bias in the width of multifractal spectrum $\Delta \alpha^{i(d)}(\gamma, L)$ with the length $L$ of the signal data and its persistency level $\gamma$:

\begin{equation}
\Delta \alpha^{c(d)}(\gamma,L)=D^{c(d)}_1 L^{-\xi^{c(d)}_1}\gamma + D^{c(d)}_0 L^{-\xi^{c(d)}_0} (1-\gamma).
\end{equation}
The same notation keeps here an obvious correspondence of indices $(c)$ or $(d)$ to the case of cumulated or differentiated data. Tables 5-6 gather values of fitted parameters present in Eq.(26) and in its extension for $95\%$ confidence level. The plots in Fig.12 collect results of semi-analytic formulas from Eqs. (21), (24), (25). The same notation as in Fig.8 applies here.
\begin{table}[!h]
\centering
\begin{tabular}{||c|c|c|c||c|c|c|c||}
\hline
$D^{c}_1$ & $\xi^{c}_1$ & $D^{c}_0$ & $\xi^{c}_0$ & $D^{c (95\%)}_1$ & $\xi^{c (95\%)}_1$ & $D^{c (95\%)}_0$ & $\xi^{c (95\%)}_0$\\ \hline
0.663&0.093&0.833&0.093&0.745& 0.100& 0.860& 0.096\\
\hline
\end{tabular}
\caption{The results of the fit for coefficients in Eq.(26) describing multifractal FSE bias for integrated data done on ensemble of $10^2$ independent realizations.}
\label{tab5}
\end{table}

\begin{table}[!h]
\centering
\begin{tabular}{||c|c|c|c||c|c|c|c||}
\hline
$D^{d}_1$ & $\xi^{d}_1$ & $D^{d}_0$ & $\xi^{d}_0$ & $D^{d (95\%)}_1$ & $\xi^{d (95\%)}_1$ & $D^{d (95\%)}_0$ & $\xi^{d (95\%)}_0$\\ \hline
1.398 &0.261&0.940&0.186&1.534& 0.274& 0.982& 0.190\\
\hline
\end{tabular}
\caption{The results of the fit for coefficients in Eq.(26) describing multifractal FSE bias for differentiated data done on ensemble of $10^2$ independent realizations.}
\label{tab6}
\end{table}

\section{Multifractality induced by basic nonlinear transformations}
It is known that nonlinear transformations introduce an additional multifractal effect even if initial data were of fully monofractal nature \cite{Havlin,mf-finance-5}. One should exactly know the level of multifractality introduced by nonlinear operations on data because one deals with already transformed data in many practical applications - for instance as volatility in finance. Particularly, there is a big challenge to discover  the presence of multifractal properties in primary not transformed data if only transformed series revealing multifractal properties is available. This problem is discussed below. 

We shall deal in this section with just two nonlinear transformations of time series increments: $\Delta x_i\rightarrow |\Delta x_i|$ and  $\Delta x_i\rightarrow (\Delta x_i)^2$ which are fundamental in modeling volatility series. Our aim is to find a semi-analytic fit to numerically simulated results of the multifractal profile of series resulting from these transformations. This will be done as previously for monofractal data of various lengths and for diversified level of long-term memory. We extend in this point the analysis done for $\Delta x_i\rightarrow (\Delta x_i)^2$ transformation in Ref.\cite{Havlin}.

As previously, we start plotting the edges of multifractal profile $h^\pm$ for transformed data against autocorrelation parameter $\gamma$ at specified series lengths.
These relations are presented for $\Delta x_i\rightarrow |\Delta x_i|$ and $\Delta x_i\rightarrow (\Delta x_i)^2$ transformation in Figs. 13, 14 respectively for data series length $L=2^{12},\ldots,2^{20}$.
The statistical uncertainty visible in these charts is calculated on an ensemble of $5\cdot10^2$ independently generated persistent time series.
A clear distinction into two regions is noticed for both transformations and for all data lengths.
We see that in the left region, the edge values $h^\pm$ descent nonlinearly with $\gamma$, while in the right one $h^\pm(\gamma)$ values remain constant.
The crossover point between abovementioned regions shall be denoted in further analysis as $\gamma^{*\pm}$, where $\pm$ corresponds to $h^\pm(\gamma)$ profiles.

In the case of main Hurst exponent $H$ calculated for transformed data, the crossover point $\gamma_H^*$ for $\Delta x_i\to(\Delta x_i)^2$ transformation was already discussed in \cite{Havlin}, and the value $\gamma^*_H=0.5$ was obtained in there.
To determine the crossover points in case of multifractal profile we present Fig. 15 with dependence $\gamma^{*\pm}(L)$.
One can clearly see that $\gamma^{*\pm}$ values change very weakly with $L$. For $\Delta x_i\to(\Delta x_i)^2$ transformation, $\gamma^{*-}$ increases to $\gamma^{*-}\approx0.6$, while $\gamma^{*+}$ drops down to $\gamma^{*+}\approx0.42$.
A qualitatively and quantitatively similar observation is made for the magnitude transformation ($\Delta x_i\to|\Delta x_i|$) shown in Fig. 15a. The crossover value $\gamma^{*-}$ remains here the same ($\approx0.6$), while $\gamma^{*+}\approx0.5$.

To determine the exact multifractal profile spread $\Delta h(\gamma,L)$, let us turn first to the region $\gamma>\gamma^*$.
In this case, $\Delta h(\gamma>\gamma^*,L) \equiv h^-(\gamma>\gamma^{*-},L)-h^+(\gamma>\gamma^{*+},L)$ is presented for both considered transformations in Fig. 16.
These plots suggest the relation 
\begin{equation}
	\Delta h(\gamma>\gamma^*,L) = C L^{-\eta} + \zeta,
\end{equation}
confirmed for both transformations in Figs.16b,c, where $\Delta h(\gamma>\gamma^*,L)-\zeta $ is shown to be linear against
 $L$ in log-log scale.
As expected, the nonlinear transformation adds a multifractal effect which, on the contrary to FSE, does not disappear for infinite length.
Therefore, the width of multifractal profile is bound by its asymptotic value ($\zeta$) depending only on the applied nonlinear transformation.
The values of fitted parameters are gathered in Table. \ref{tab7} for both considered transformations.
The fitting result of Eq. (27) as well as its 95\% confidence level arising from statistics is presented in Fig. 16a (see the top blue curves).

\begin{table}[!h]
\centering
\begin{tabular}{||c||c|c|c||c|c|c||}
\hline
Transformation & $C$ & $\eta$ & $\zeta$ & $C^{95\%}$ & $\eta^{95\%}$ & $\zeta^{95\%}$ \\\hline
$\Delta x_i\rightarrow |\Delta x_i|$   & 1.699 & 0.376 & 0.040 & 2.012 & 0.374 & 0.043 \\\hline
$\Delta x_i\rightarrow (\Delta x_i)^2$ & 16.404 & 0.643 & 0.118 & 17.031 & 0.632 & 0.126\\
\hline
\end{tabular}
\caption{Values of parameters found by fitting Eq.(27) altogether with their 95\% confidence levels resulting from statistical and fitting uncertainties.}
\label{tab7}
\end{table}

The parameter $\zeta$ can be interpreted as the actual amount of multifractal effect introduced solely by nonlinear transformations.
One can clearly see that scaling parameter $\eta$, as well as the asymptotic value $\zeta$ of $\Delta h(\gamma,L)$  at the limit $L\to\infty$, has higher values for $\Delta x_i\to(\Delta x_i)^2$ transformation.
It proves therefore that the latter transformation introduces stronger observed multifractal effect.

One can make also a trial to describe semi-analytically the $\gamma<\gamma^*$ region.
We have found from series of plots in Figs. 13, 14, that in this region the $q^{'}$-exponential formula is the best candidate for fitting the nonlinear descent $h^\pm(\gamma<\gamma^*)$
\begin{equation}
	h^{\pm}(\gamma<\gamma^{*\pm},L) = h^{\pm}(\gamma>\gamma^{*\pm},L)\exp_{q'}\big\{-A^\pm(L)(\gamma-\gamma^{*\pm})\big\}
	\label{qexp}
\end{equation}
where \cite{qexp}
\begin{equation}
	\exp_{q^{'}}(-x)=\left[1-(1-q^{'})x\right]^{\frac{1}{1-q^{'}}}
\end{equation}
The value of $q^{'}$ parameter\footnote{$q^{'}$ should not be confused with $q$ appearing in MF-DFA method; the similar notation was used in sake of traditionally grounded notation} was found $q^{'}\cong 1.6$ for both considered transformations by minimizing the mean squared error of fit. The values $A^\pm(L)$ are responsible for the shape of $q^{'}$-exponential decay of multifractal profile edges $h^\pm$ with $\gamma$, while $\gamma^{*\pm}$ are the actual crossover points between regions of nonlinear and constant behaviors of $h^{\pm}$.
The crossover values are gathered in Table 8.

Figs. 17 and 18 prove that $A^{\pm}(L)$ can be taken constant for both transformations within statistical uncertainty of fit to Eq.(28).
Thus the $\Delta h$ dependence on data length for $\gamma<\gamma^*$ is fully determined by the spread $\Delta h(\gamma>\gamma^*,L)$.
The results of fit to central values $A^\pm$, as well as the corresponding values at 95\% confidence level\footnote{by this fit we mean independent fitting of Eq.(\ref{qexp}) to numerically estimated values $h^\pm(\gamma<\gamma^{*\pm},L)$ adjusted by statistical uncertainty coming from the ensemble of $5\cdot10^2$ independent simulations of time series} are gathered in Table 9.
\begin{table}[!h]
\centering
\begin{tabular}{||c||c|c||c|c||}
\hline
Transformation & $\gamma^{*+}(L)$ & $\gamma^{*-}(L)$ & $\gamma^{*+95\%}(L)$ & $\gamma^{*-95\%}(L)$ \\\hline
$\Delta x_i\rightarrow |\Delta x_i|$   & 0.52 & 0.63 & 0.53 & 0.63 \\
\hline
$\Delta x_i\rightarrow (\Delta x_i)^2$ & 0.40 & 0.61 & 0.42 & 0.61 \\
\hline
\end{tabular}
\caption{Results of fit to $\gamma^{*\pm}(L)$ parameters in $q$-exponential formula (Eq. (\ref{qexp})) altogether with the 95\% confidence level values. The uncertainty of fit was $\pm 0.01$.}
\label{tab9}
\end{table}
%
\begin{table}[!h]
\centering
\begin{tabular}{||c||c|c||c|c||}
\hline
Transformation & $A^+$ & $A^-$   & $A^{+95\%}$ & $A^{-95\%}$ \\\hline
$\Delta x_i\rightarrow |\Delta x_i|$   & 0.95 & 0.83 & 0.94 & 0.84 \\
\hline
$\Delta x_i\rightarrow (\Delta x_i)^2$ & 0.87 & 1.02 & 0.84 & 1.03 \\
\hline
\end{tabular}
\caption{Results of fit to $A^\pm(L)$ parameter in $q$-exponential formula (Eq. (\ref{qexp})) altogether with the 95\% confidence level thresholds.}
\label{tab8}
\end{table}

The summary given by Eqs. (27) and (28) for the dependence of the $\Delta h$ spread on persistency level $\gamma$ for various lengths of data is presented in Fig. 19. Two different behaviors indicated for $\gamma$ above and below crossover value are noticeable. The transition area is visible due to the fact that two crossover values do not coincide  ($\gamma^{*+}<\gamma^{*-}$).

\section{Concluding remarks}
In the first part of this paper we have shown qualitatively and quantitatively how the observed multifractal effects arise in monofractal series of finite uncorrelated or persistent data. This kind of multifractality, called by us multifractal bias of finite size effect, should be clearly distinguished from the 'true multifractality' caused by memory effects dependent on the time scale and thus related to different scaling properties of data at various time scales. The quantitative findings presented here confirm and push further the study of the generalized Hurst exponent spread suggested recently in \cite{schu-kantelhardt}, to be not indicative for multifractality if $\Delta h \sim 0.2$ or $\Delta \alpha \sim 0.3$ are obtained.
Although, found results refer to clearly synthetic monofractal series, they are important in practical applications, where in general,  multifractal data are observed. 
Results of the same methodology are affected in a similar manner for mono- and multifractal data and a similar level of the bias will appear in all multifractal data as well.

We considered two methods of description for multifractality, i.e. the generalized Hurst
exponent approach and the multifractal spectrum analysis based on the H\H{o}lder exponent. In both cases the multifractal threshold of finite data was estimated by the spread of generalized Hurst exponent or by the width of multifractal spectrum. The threshold was shown to grow linearly with
autocorrelation level in time series and to decay according to the power-law with series length. We
have estimated numerically the level of such apparent multifractal effect  and we captured it in simple semi-analytical formulas.

Additionally, in the second part of the article, we have discussed the multifractality revealed by the nonlinearly transformed monofractal data.
In such a case, the finite size effects amplified by long-range autocorrelations in primary data are accompanied also by the fluctuation clusters arising from such nonlinear transformation of time series. This last effect changes the scaling properties of various fluctuation sizes and does not vanish in the limit of infinite data length. One must also keep in mind the influence introduced by nonlinear transformations, especially when only transformed data are searched and the goal are  multifractal properties of the primary data series. The result of multifractal effect can be significantly dimmed in there.

In all cases the ready to use formulas describing amount of multifractal FSE threshold or a multifractal modification of time series properties after some simple nonlinear transformations  were given.
We provided also results for statistical confidence at 95\% level   for all formulas, what enables the reliable estimation when they are used in practical applications.
We believe these formulas are general enough to be applied in any area of science when one is questioning the multiscale properties of data.

Eventually, the predicted FSE multifractal bias should be compared with examples of
real multifractal records. There is a common agreement that
multifractality is a characteristic feature for financial markets, so we may first look at index series.
The multifractal features for various price indices were taken from the general study of Ref. \cite{zunino}. By applying formulas from Eq.(17) to parameters of real index series revealed in \cite{zunino}, we easy find that  multifractality comes indeed as a result of scaling properties changing with the time scale for many world markets. However, there are markets where the observed multifractal features are generated mainly (e.g. Philippines, Taiwan, Germany) or even entirely (Ireland)
by the finite size effects. In the case of Philippines, Taiwan, Thailand, Germany, Spain and
Greece almost 80\% of  the observed multifractality in price indices is caused by multifractal FSE bias \footnote{see the conference paper \cite{FENS} for detailed study}. 
In Ref. \cite{zhou} it is reported that FSE (achieved for finite size surrogate data) for DJIA time series counts for $\Delta h=0.22\pm 0.04 $ at $L\sim 30000$, which  agrees with the result of  Eq. (17) predicting $\Delta h=0.23$ at $95\%$ confidence level.

Regarding other kinds of real data, e.g. in seismology, meteorology or geophysics, an importance of multifractal FSE bias is also indisputable.
The seismological records from Refs. \cite{MF-seismology_1,MF-seismology_2} suggest multifractal spread of generalized Hurst exponent  $\Delta h\approx0.53$ for $L\sim2^{13}$ at $H\sim0.9$ (see Fig.3 in \cite{MF-seismology_1}) or the spectrum width $\Delta \alpha\approx0.70$ (see Fig.5 in \cite{MF-seismology_2}),
while the calculated bias from Eqs.(17) and (20) counts for more than $40\%$ of this value in case of $\Delta\alpha$ and $\simeq 30\%$ in case of $\Delta h$.

Many records in meteorology and geophysics reveal high nonstationarity, so that authors focus on much narrower range of $q$ ($-5<q<5$) then used in this article.
The quoted multifractal quantitative features for records in meteorology (see \cite{MF-meteo}) are reported very low for	humidity, temperature, solar radiation and maximum squall ($\Delta \alpha\sim 0.02\div0.07$) and more significant for precipitation, wind speed and atmospheric pressure ($\Delta\alpha\sim 0.20\div0.28$, $L\approx2^{12}$, $H\approx0.6\div0.7$).
However, to make a  proper comparison with the FSE multifractal bias in such a case, one has to examine how formulas obtained in this article change when lower moment $q$ is used for evaluation. The light on such problem is shed in coming publication \cite{arxiv-Qdependence}.

Thus, the multifractal bias (or apparent multifractality) may play significant role in the observed multifractal spread, especially when short and persistent time series are considered.
In many cases, it aggravates the separation between the main multiscaling phenomenon and the background (bias) of not multiscaling origin at all.
Therefore, one should be especially careful drawing far reaching
conclusions from the multifractal analysis based on interpretation of observed $\Delta h$ or $\Delta \alpha$ values.

\begin{figure}[p]
\includegraphics[width=11truecm]{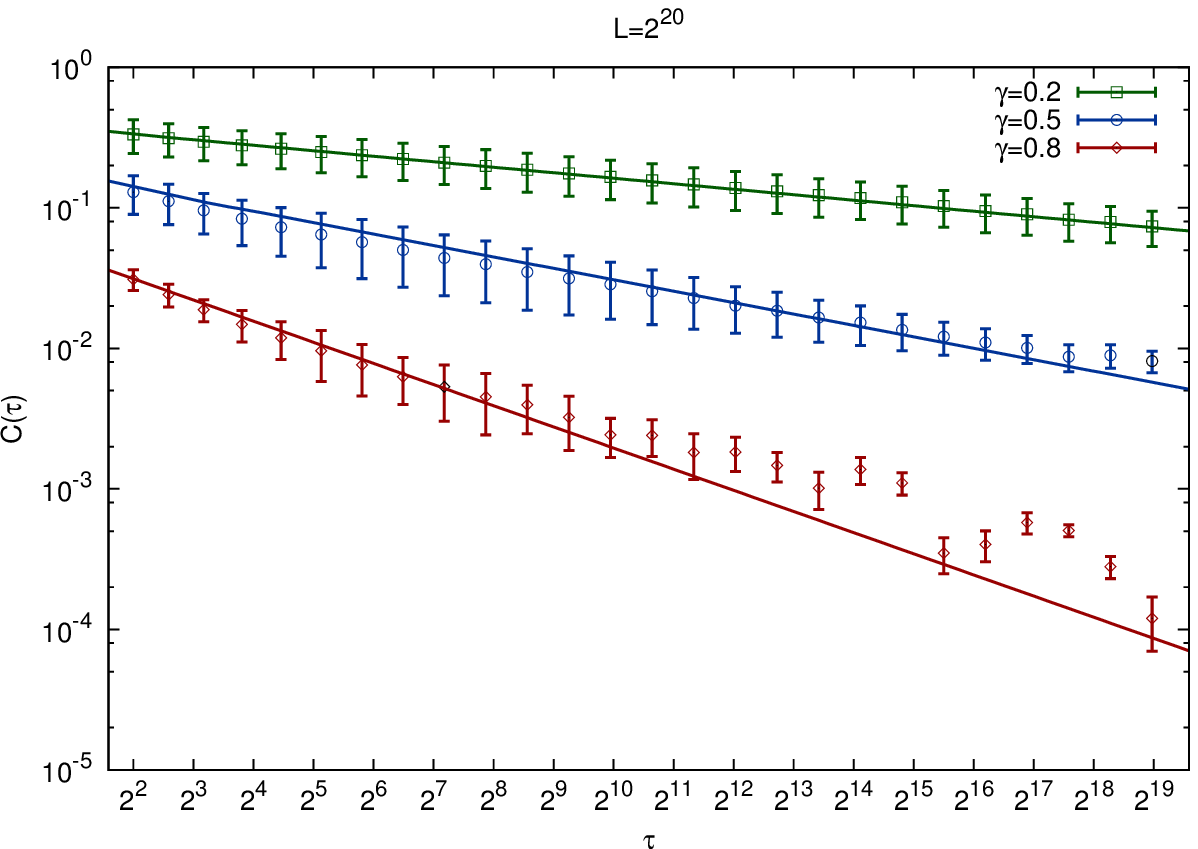}
\caption{Efficiency of FFM for replication of autocorrelation properties in time series. The examples for input values $\gamma=0.2$, $0.5$, $0.8$ are shown in log-log scale for the generated data of length  $L=2^{20}$. The lines present the fit to the desired power law dependence of Eq.(1), while error-bars show $1\sigma$ standard deviation following from the considered statistics of $10^2$ independent realizations. The output $\gamma$ values from the fit are found $\gamma_{out}=0.203 (\pm 0.009)$, $0.498 (\pm 0.012)$, $0.782 (\pm 0.054)$ respectively.}
\label{c-l}
\end{figure}
\clearpage

\begin{figure}[p]
\includegraphics[width=16truecm]{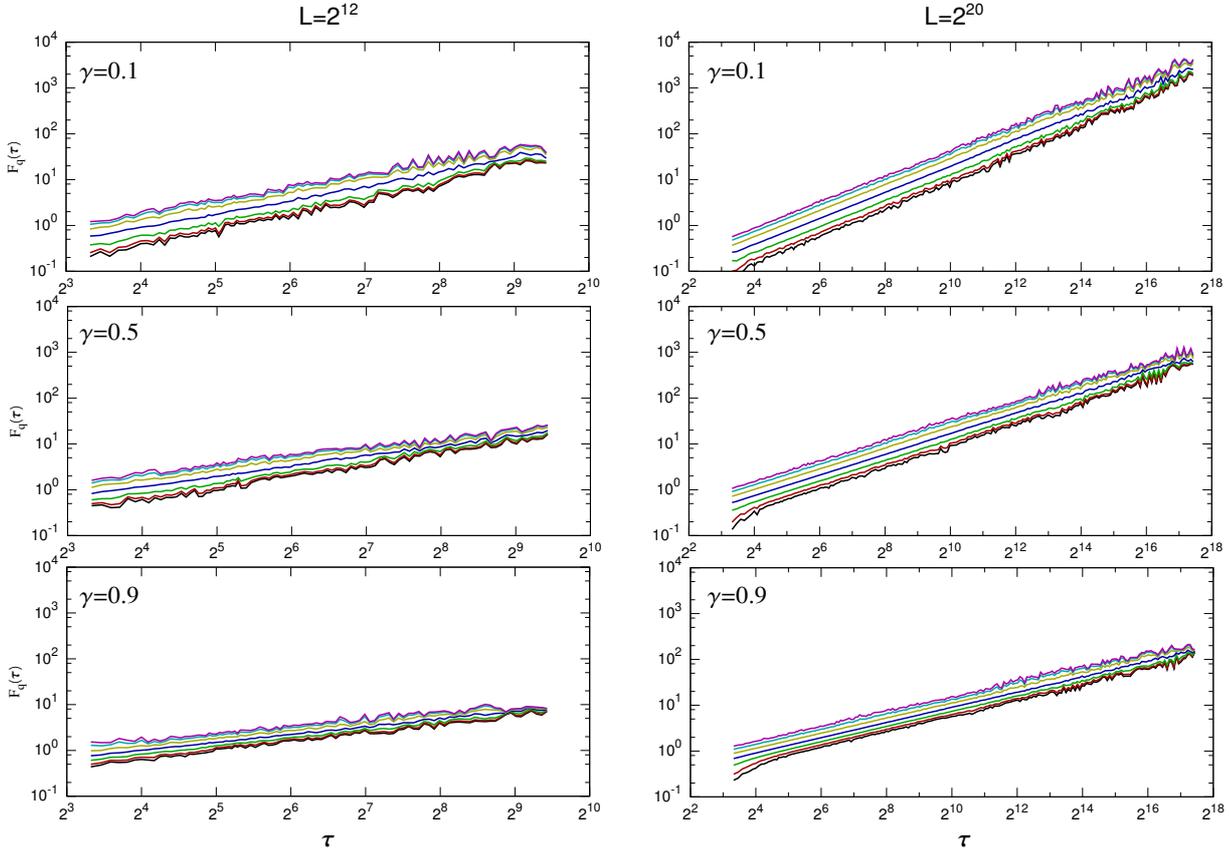}
\caption{Scaling of $q$-deformed fluctuations within MF-DFA. Results are presented for two different lengths of time series $L=2^{12}$, $2^{20}$, three autocorrelation parameters $\gamma=0.1$, $0.5$, $0.9$ and $q=-15,-10,-5,0,+5,+10,+15$ (from bottom to top). All plots confirm the proposed scaling range from $\tau=20$ till $\tau=L/4$.}
\label{Fq-tau}
\end{figure}
\clearpage

\begin{figure}[p]
\subfloat[][]{
\includegraphics[width=8truecm]{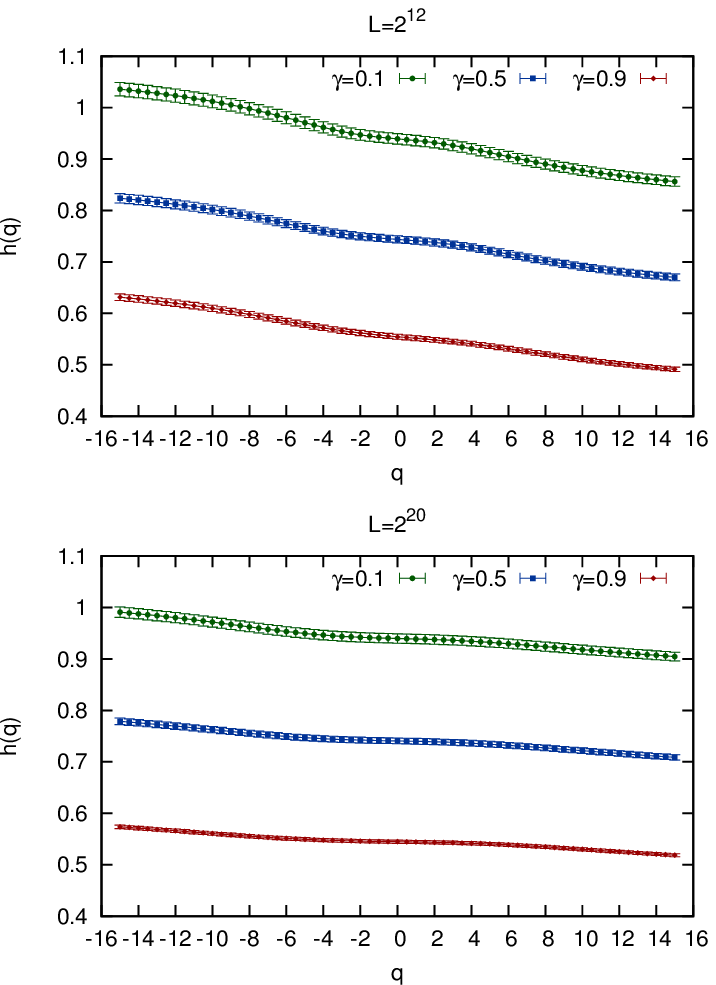}
}
\subfloat[][]{
\includegraphics[width=8truecm]{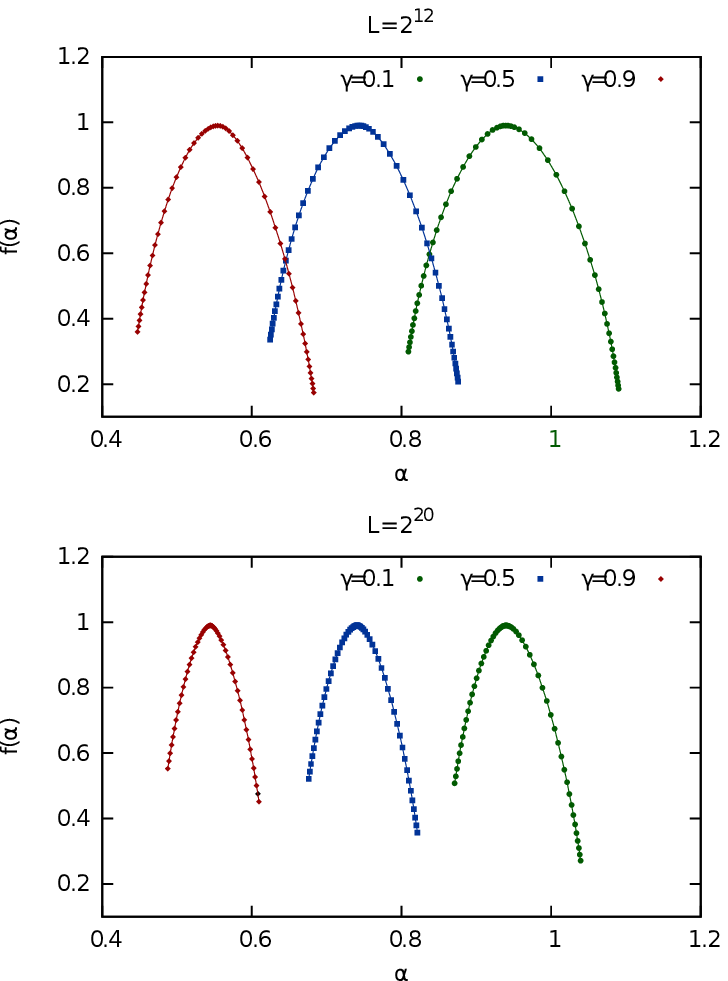}
}
\caption{Generalized Hurst exponent (a) and singularity spectrum (b) for monofractal signals generated within FFM with various autocorrelation properties. Two cases, for $L=2^{12}$ and $L=2^{20}$ are shown with different autocorrelation levels. Error bars correspond to statistics of $10^2$ generated series.}
\label{h_sh-q}
\end{figure}
\clearpage

\begin{figure}[p]
\subfloat[][]{
\includegraphics[width=8truecm]{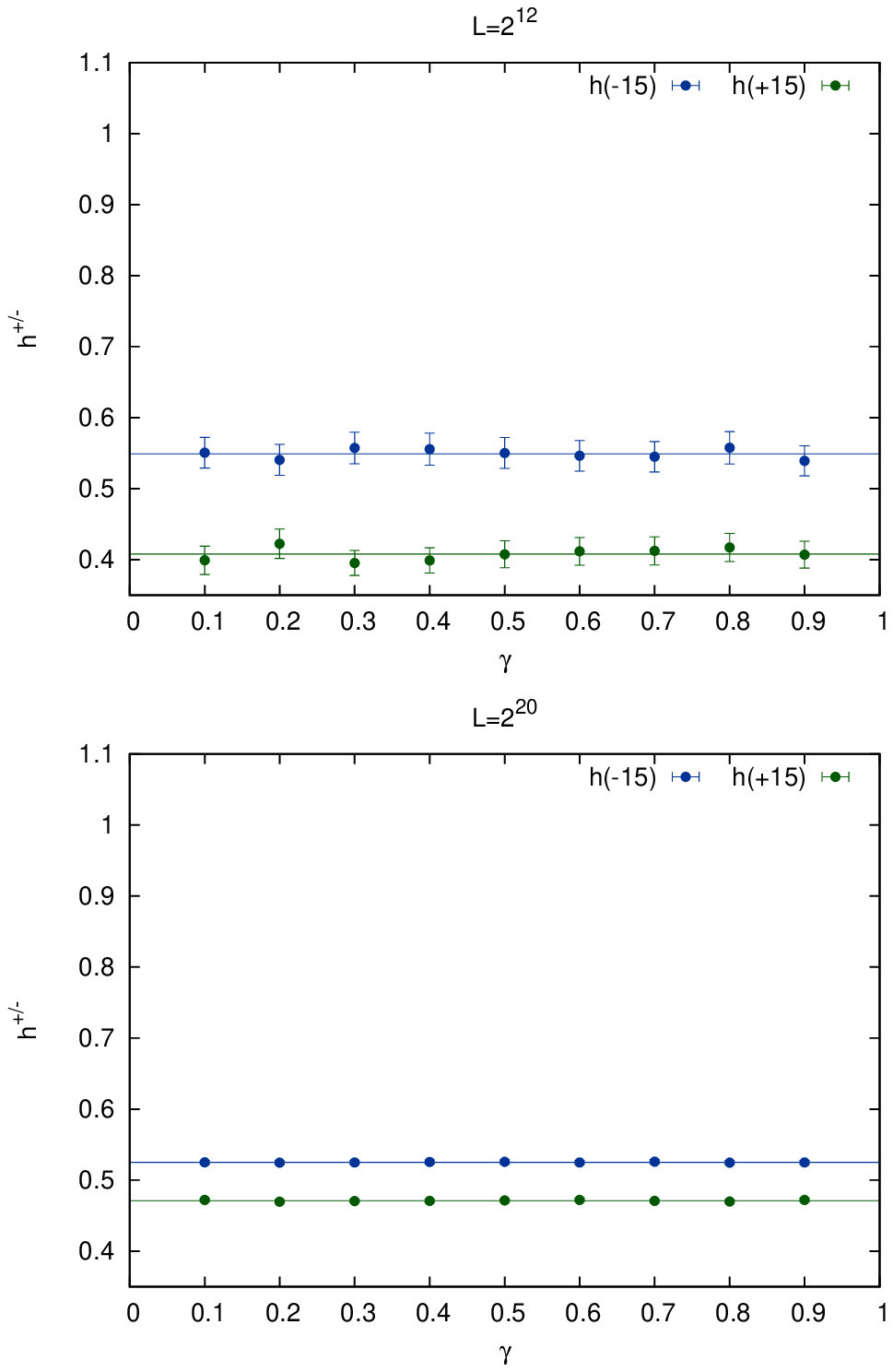}
}
\subfloat[][]{
\includegraphics[width=8truecm]{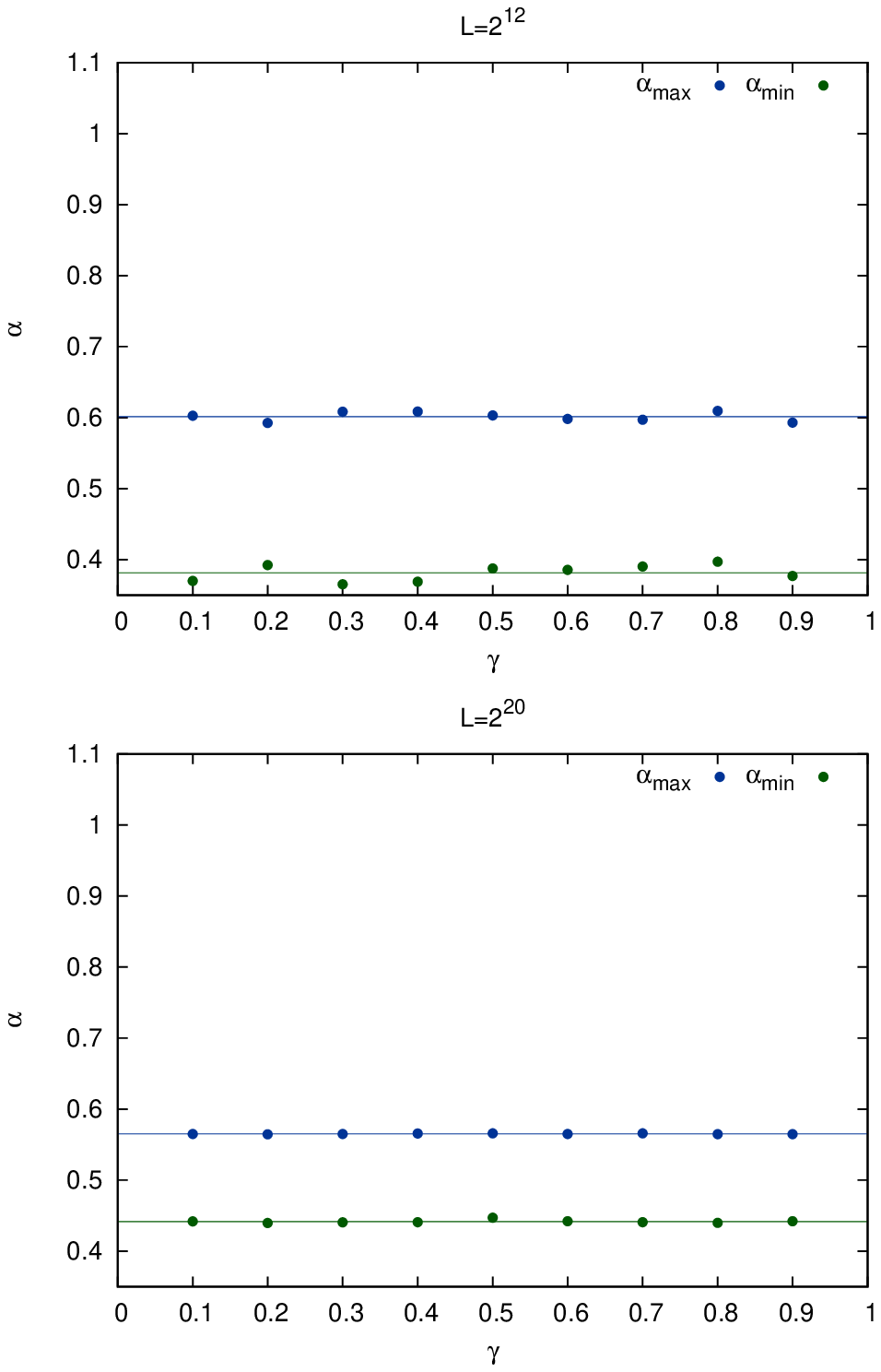}
}
\caption{Edge values of the generalized Hurst exponents $h(q)$ (a) and H\"older parameter $\alpha$ (b) for two different lengths of time series $L=2^{12}$, $2^{20}$ constructed with long memory present ($\gamma<1$) and then shuffled to kill this memory. Dependence on the data length is readable.}
\label{h_edge_sh-gamma}
\end{figure}
\clearpage

\begin{figure}[p]
\subfloat[][]{
\includegraphics[width=8truecm]{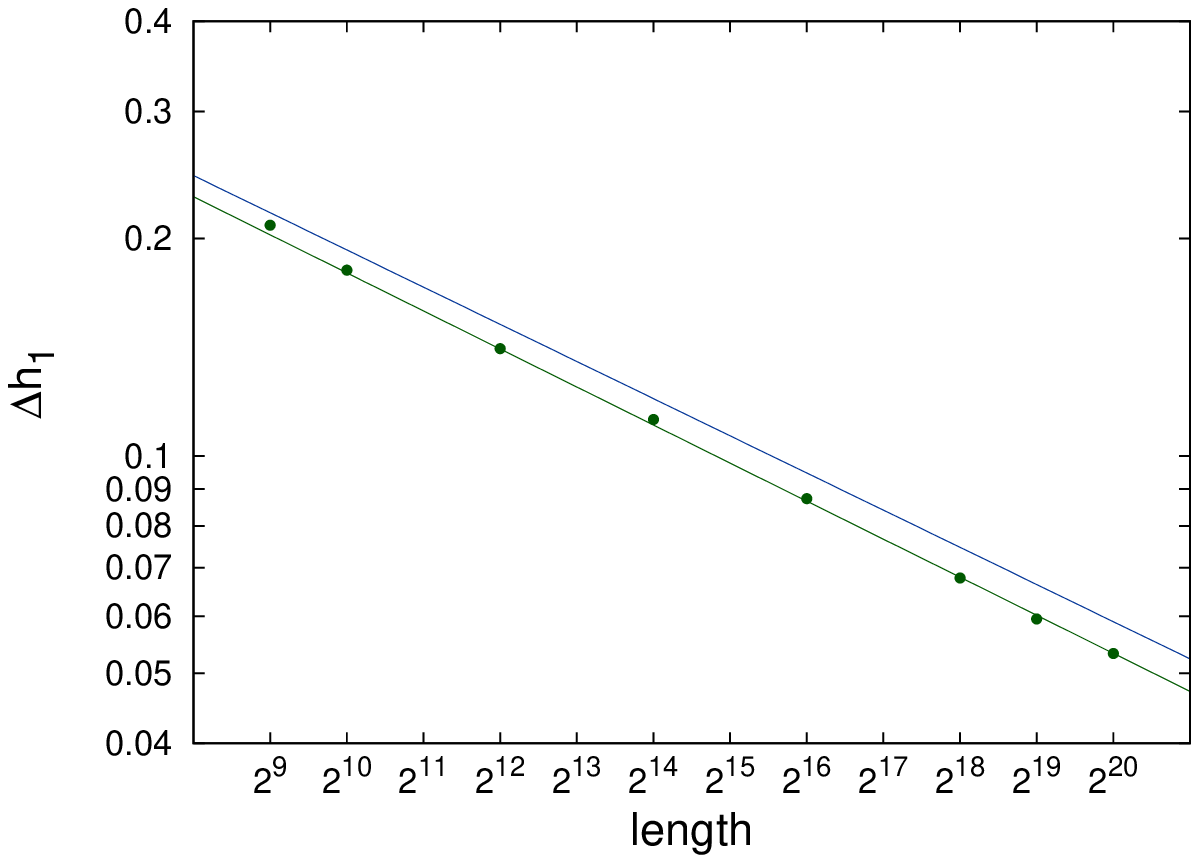}
}
\subfloat[][]{
\includegraphics[width=8truecm]{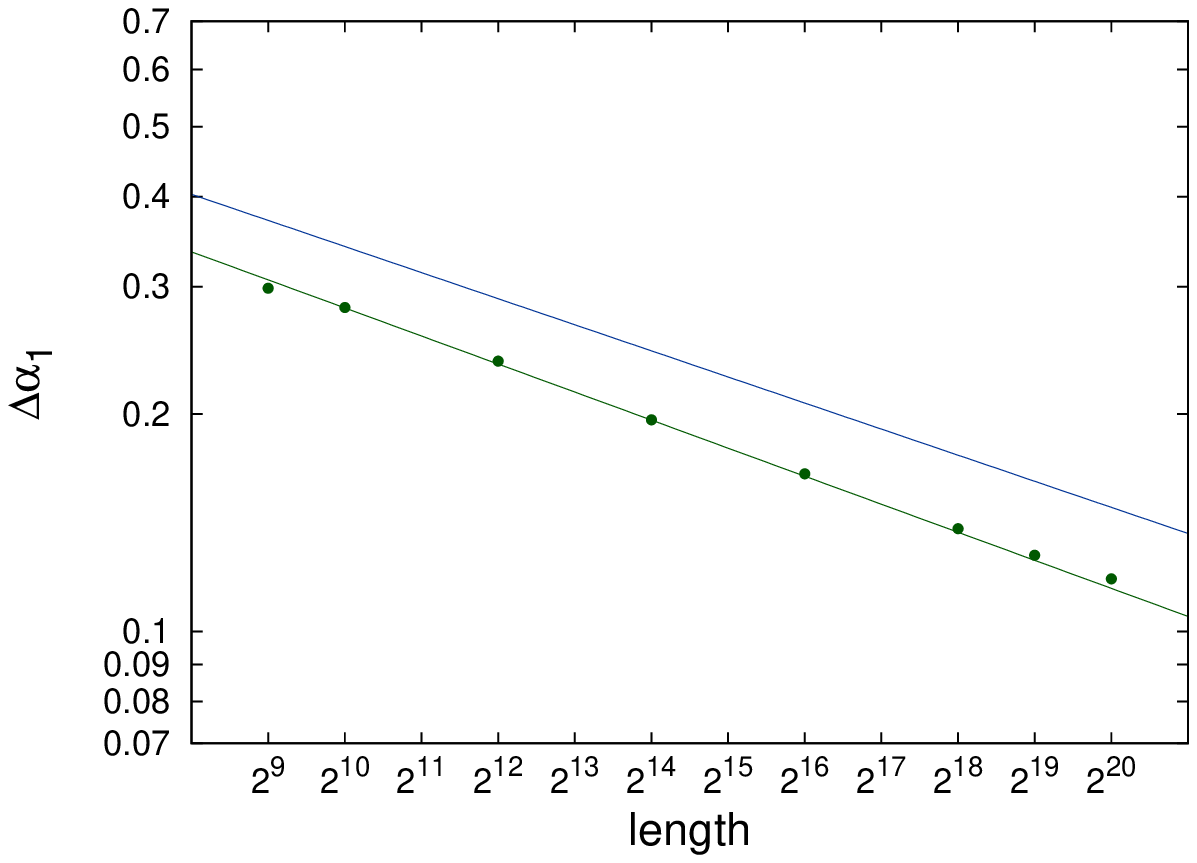}
}
\caption{Spread $\Delta h$ of generalized Hurst exponent (a) and spread $\Delta\alpha$ of H\"older parameter (b) versus length of data drawn in logarithmic scale for the signal with no memory. Power-law dependence between  $\Delta h$ ($\Delta\alpha$ in (b) case) and the data length is visible. Results of the fit are gathered in Tables 1 and 2 for central values as well as for their 95\% confidence level. The latter fit is marked as the blue top line.}
\label{h_sh-q}
\end{figure}
\clearpage

\begin{figure}[p]
\subfloat[][]{
\includegraphics[width=8truecm]{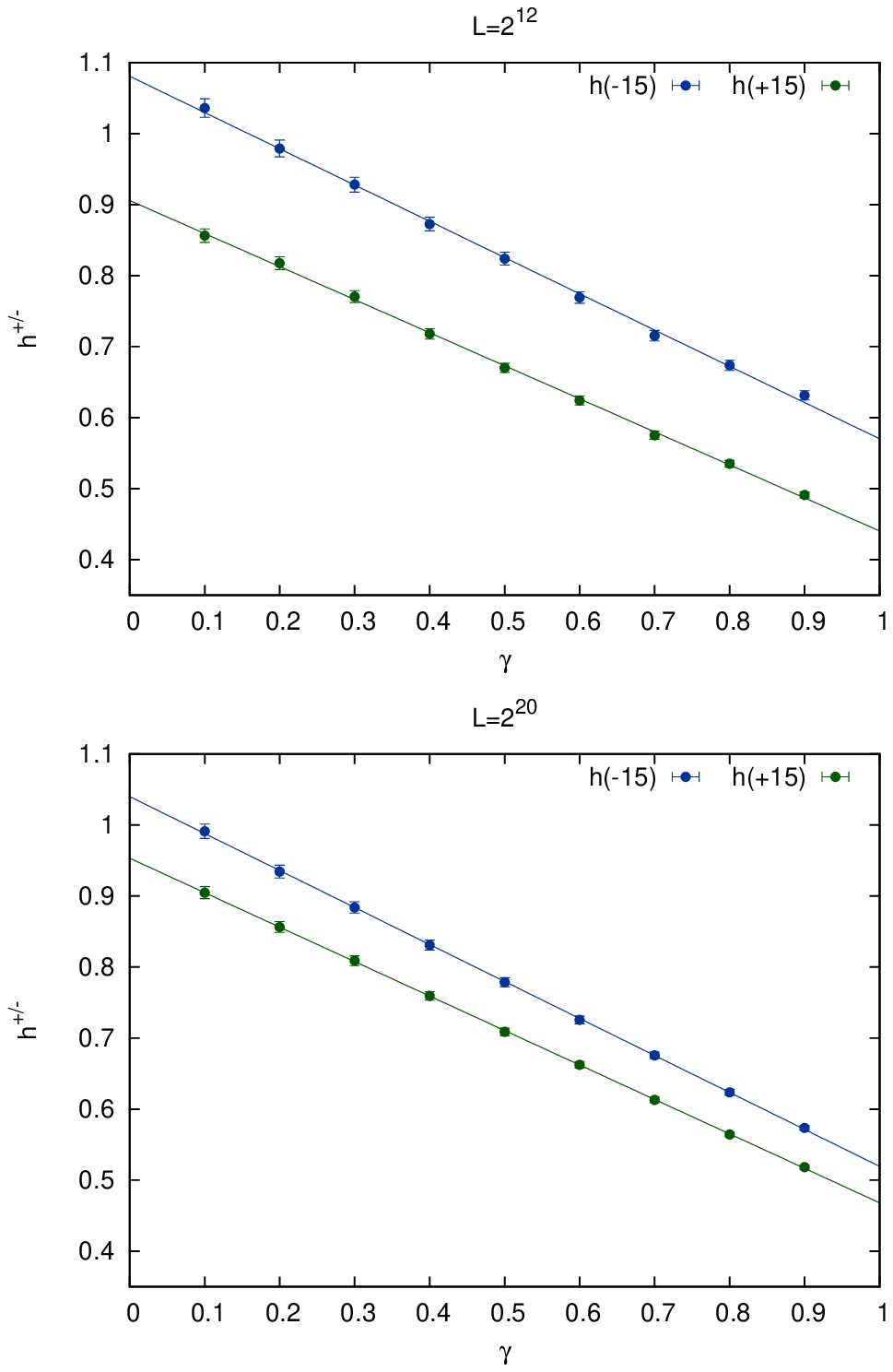}
}
\subfloat[][]{
\includegraphics[width=8truecm]{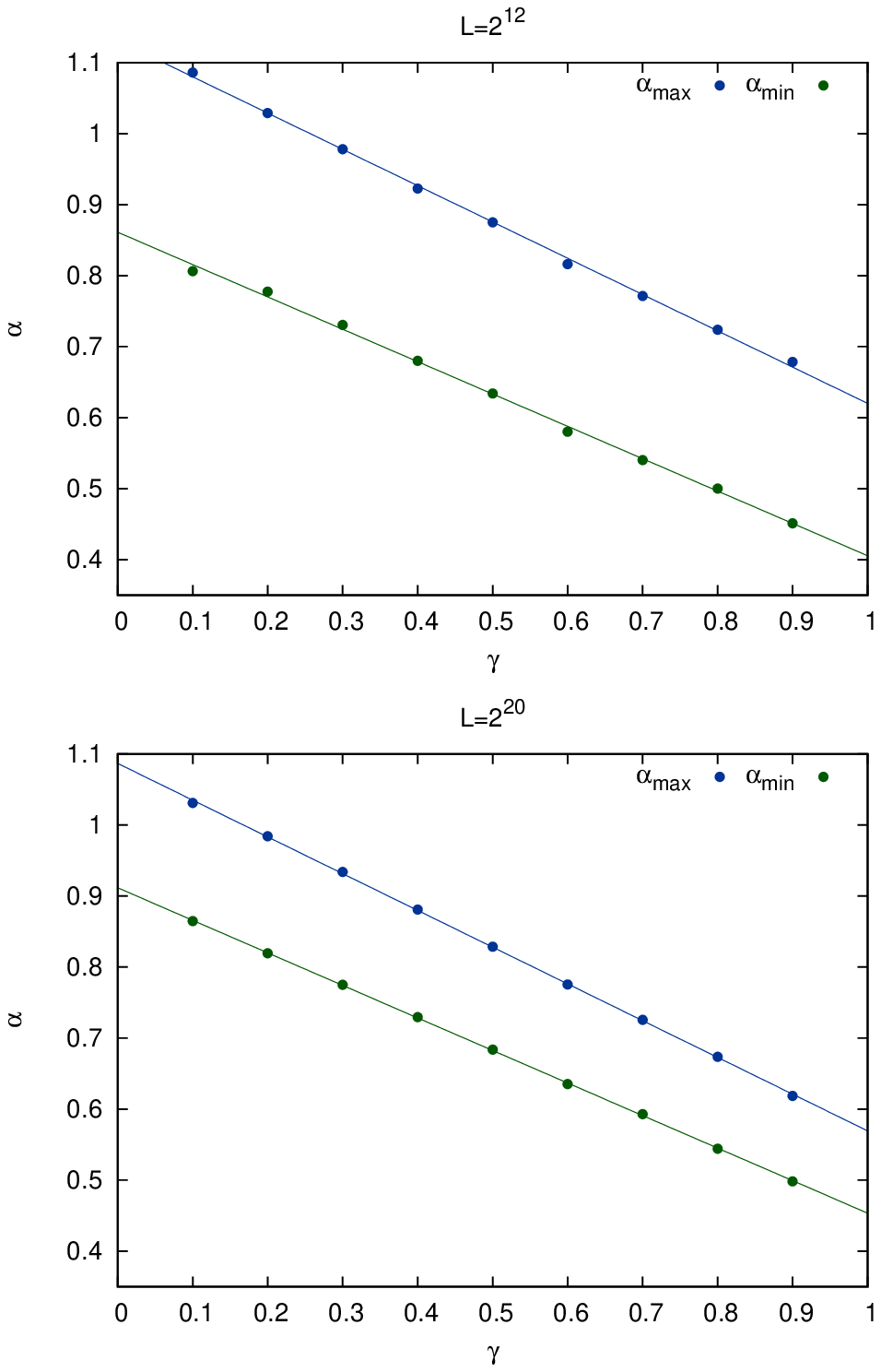}
}
\caption{Edge values of generalized Hurst exponent (a) and H\"older parameter (b) for series with long-term memory. Figures show the linear dependence between the edge values $h^{\pm}\equiv h(\pm 15)$ and $\alpha_{min/max}$ on $\gamma$ exponent. Extrapolation of fitted lines to the point $\gamma=0$ are interpreted as the edge values for fully autocorrelated signal ($C(\tau)\rightarrow 1,\ \forall\tau$)}
\label{h_sh-q}
\end{figure}
\clearpage

\begin{figure}[p]
\subfloat[][]{
\includegraphics[width=8truecm]{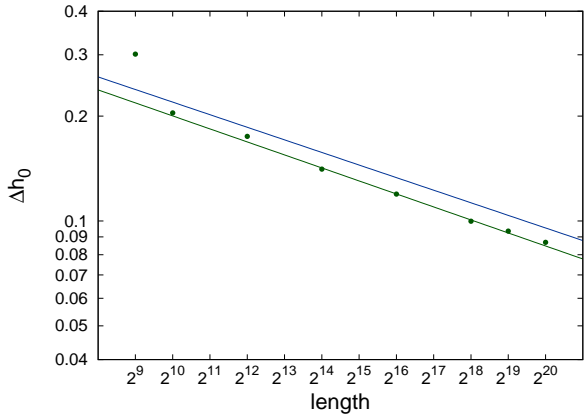}
}
\subfloat[][]{
\includegraphics[width=8truecm]{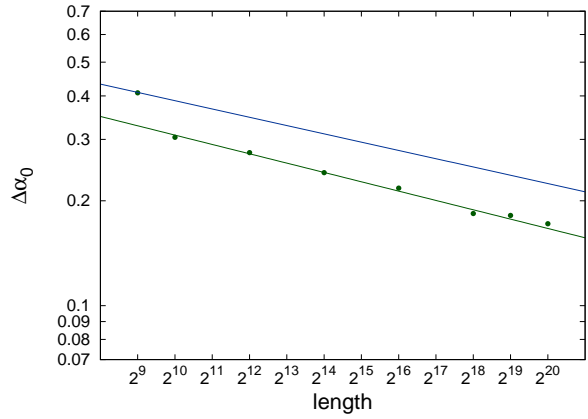}
}
\caption{Spread $\Delta h_0$ of generalized Hurst exponent (a) and $\Delta\alpha_0$ of H\"older parameter for fully autocorrelated time series ($\gamma=0$) versus the length of data. Green line presents the power-law  fit in log scale and the blue line corresponds to $95\%$ confidence level resulting from statistics. Fitted parameters are gathered in Table 1 for (a) and Table 2 for (b). Data point corresponding to $L=2^9$ has been removed from the fit due to insufficient statistics for so short signal leading to huge uncertainty in the estimation of generalized Hurst exponents within MF-DFA.}
\label{h_sh-q}
\end{figure}
\clearpage

\begin{figure}[p]
\includegraphics[width=15truecm]{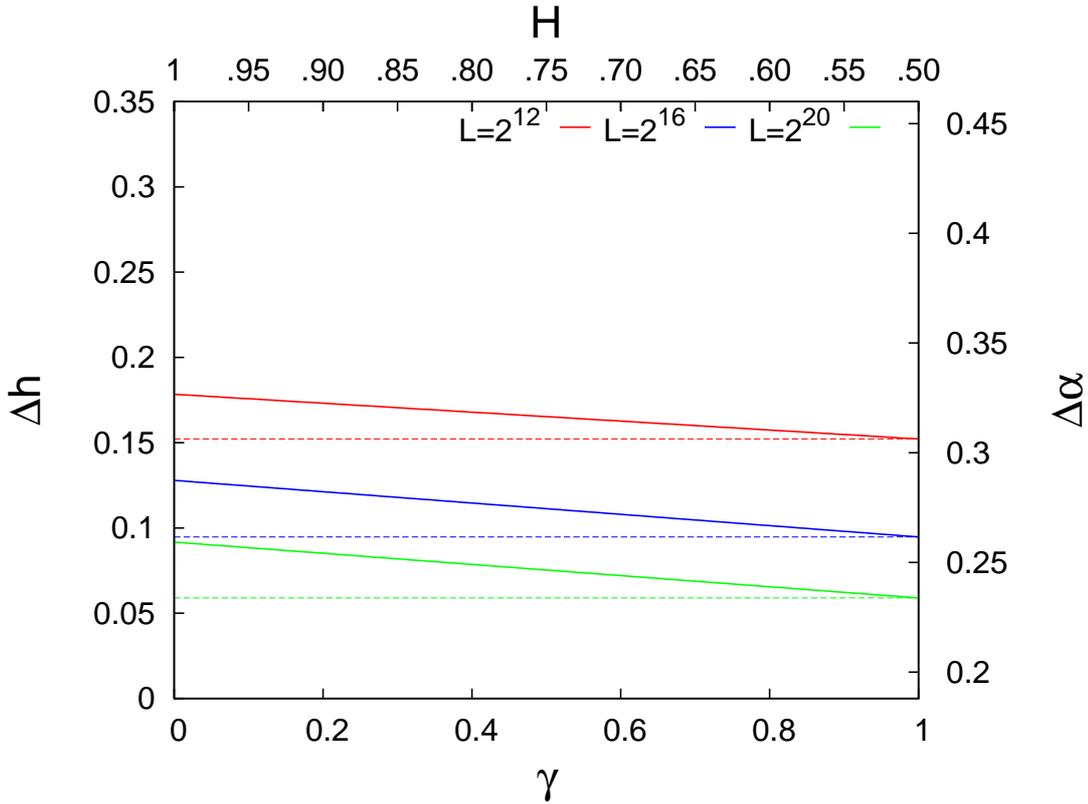}
\caption{Examples of $\Delta h$ and $\Delta \alpha$ FSE multifractal thresholds for variety of data lengths $L= 2^{12}$, $L= 2^{16}$ and $L= 2^{20}$. The continuous sloped lines describe the minimal threshold for multiscaling  to be observed at $95\%$ confidence level for series with autocorrelation exponent $\gamma$. The respective values of Hurst exponent are marked on top axis. The dotted horizontal lines indicate the corresponding multifractal FSE threshold resulting only from finite data length not affected by possible autocorrelations in data.}
\label{h_sh-q}
\end{figure} 
\clearpage

\begin{figure}[p]
\centering
$L=2^{12}$\\
\vspace*{-1em}
\subfloat[][]{
\includegraphics[width=8truecm]{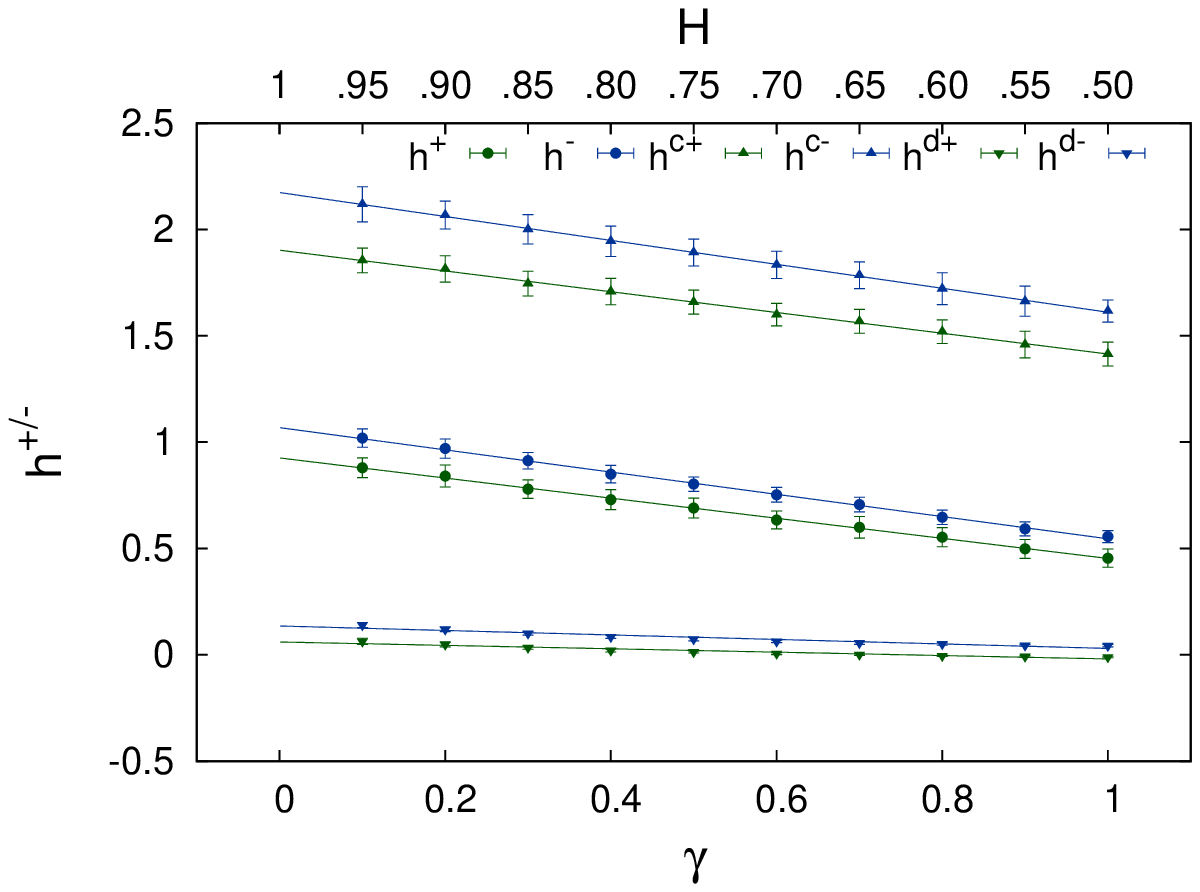}
}
\subfloat[][]{
\includegraphics[width=8truecm]{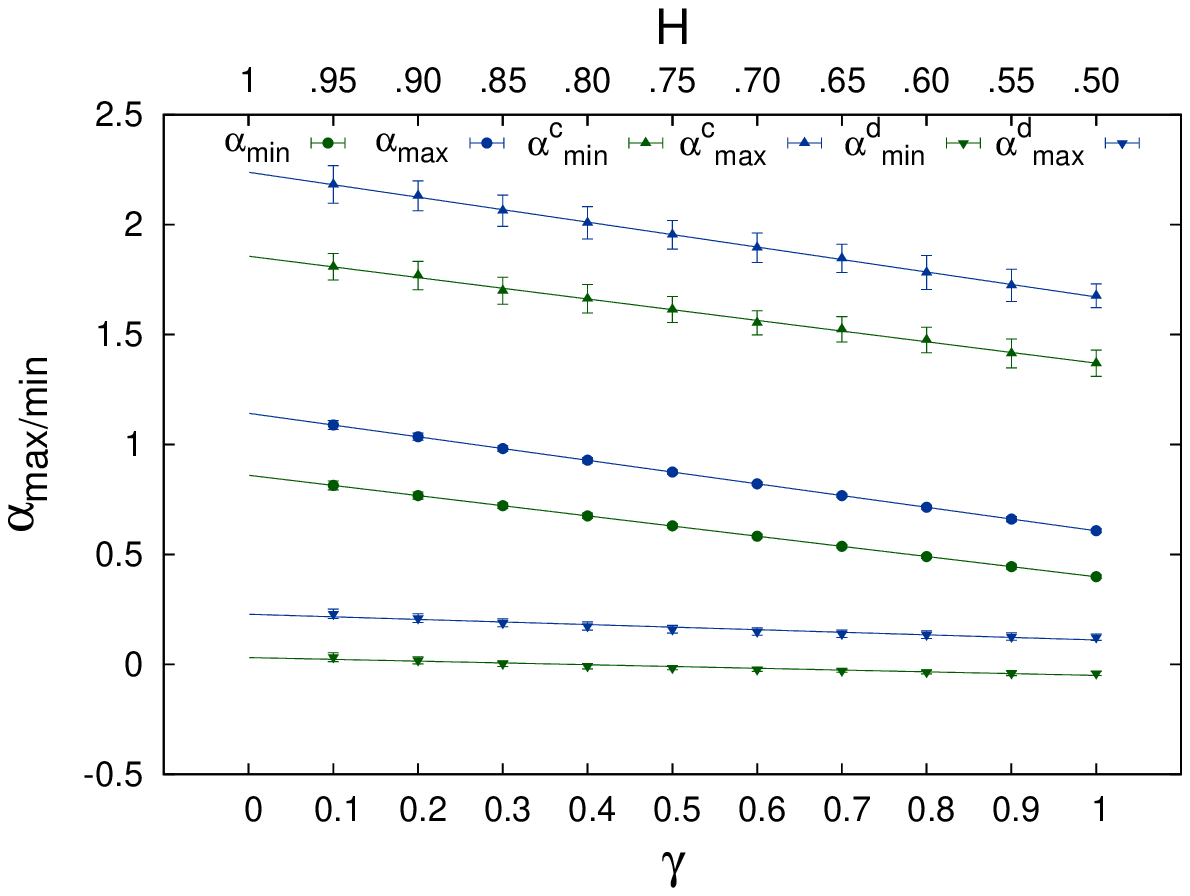}
}

$L=2^{16}$\\
\vspace*{-1em}
\subfloat[][]{
\includegraphics[width=8truecm]{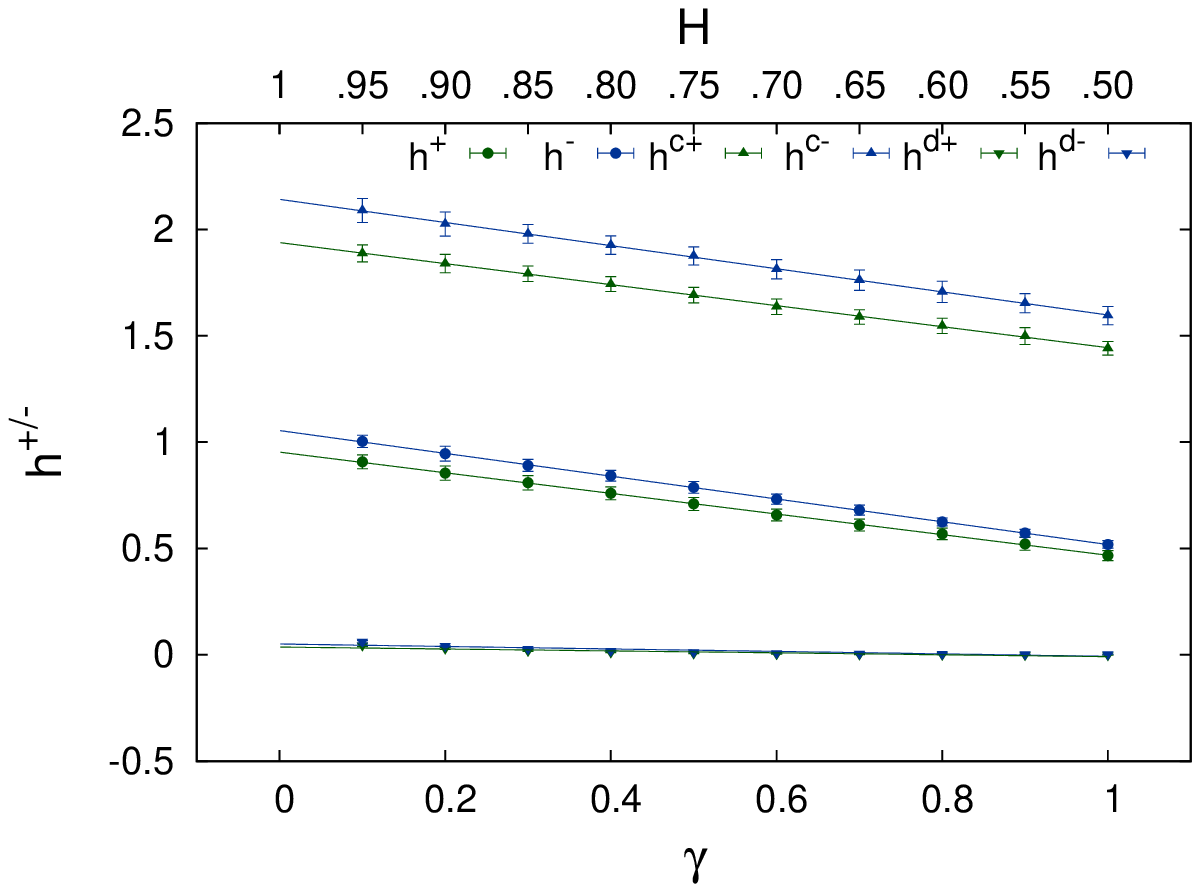}
}
\subfloat[][]{
\includegraphics[width=8truecm]{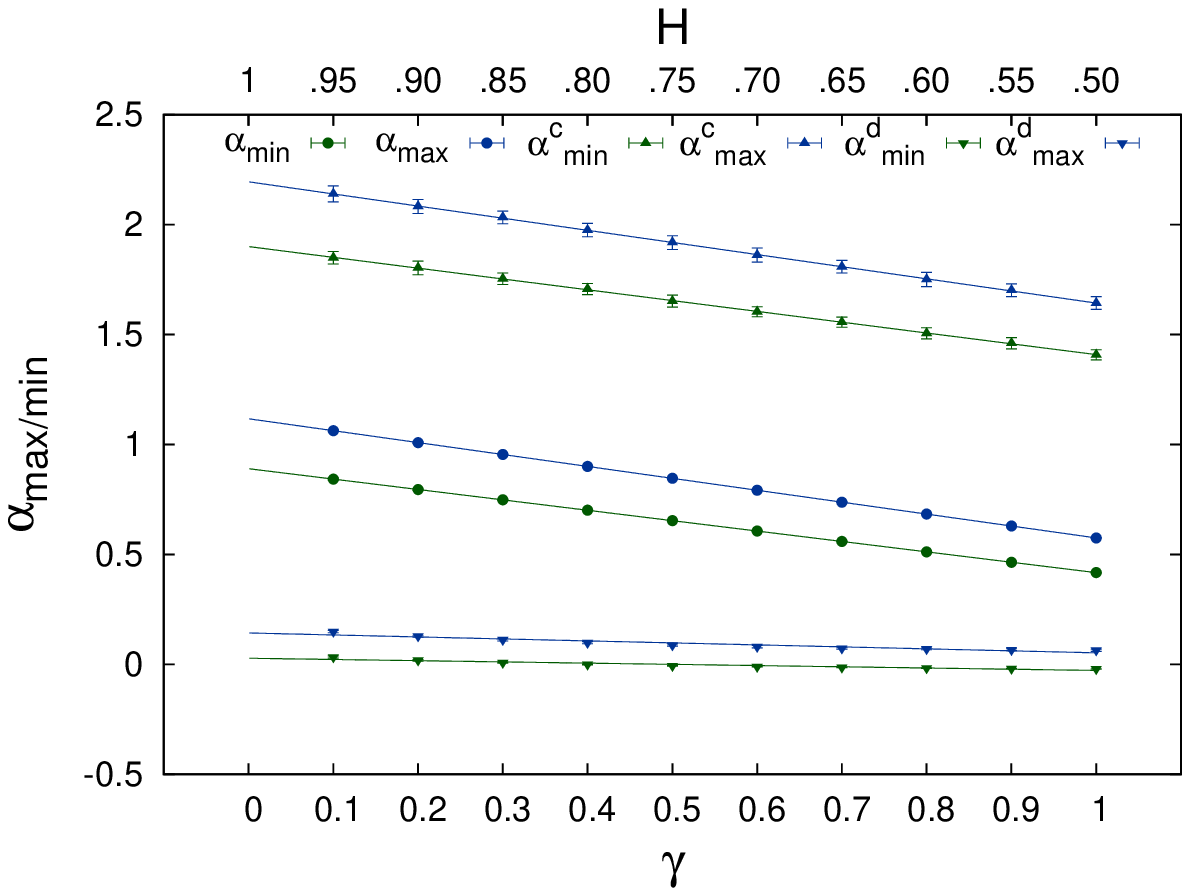}
}

$L=2^{20}$\\
\vspace*{-1em}
\subfloat[][]{
\includegraphics[width=8truecm]{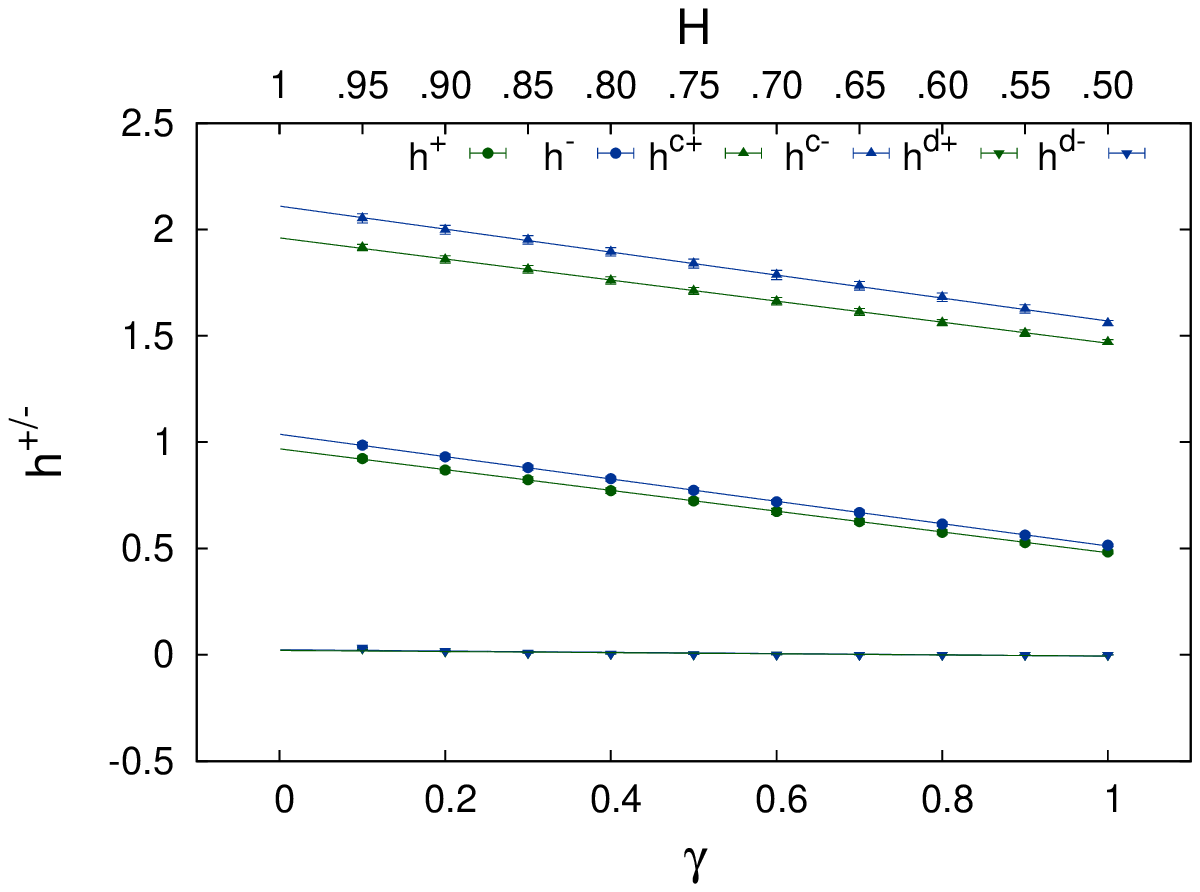}
}
\subfloat[][]{
\includegraphics[width=8truecm]{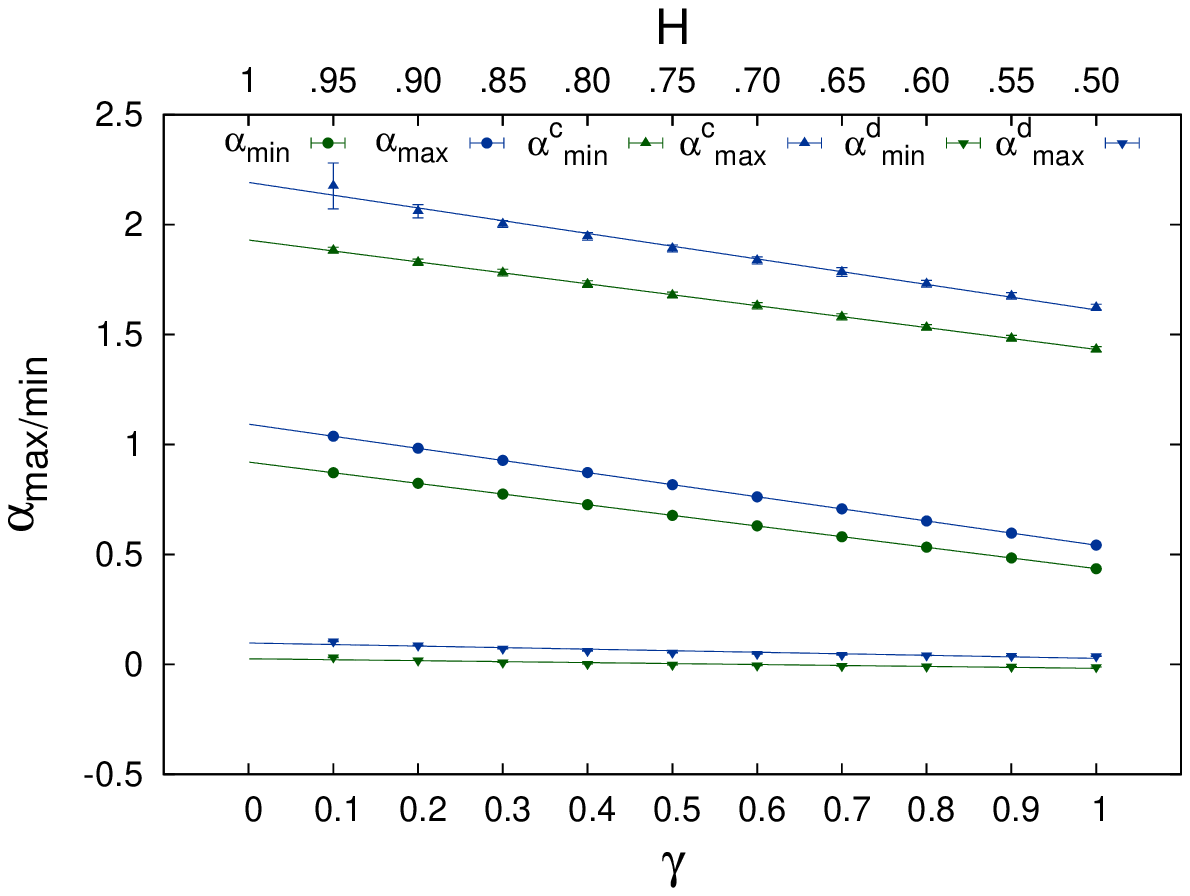}
}
\caption{Edge values of generalized Hurst exponent (a) and H\"older parameter (b) for primary, integrated and differentiated series with long-term memory of lengths $L=2^{12},2^{16},2^{20}$. Figures show the linear dependence between the edge values $h^{\pm}\equiv h(\pm 15)$ and $\alpha_{min/max}$ on $\gamma$ exponent. Extrapolation of fitted lines to the point $\gamma=0$ are interpreted as the edge values for fully autocorrelated signal ($C(\tau)\rightarrow 1,\ \forall\tau$)}
\label{}
\end{figure}
\clearpage

\begin{figure}[p]
\subfloat[][]{
\includegraphics[width=8truecm]{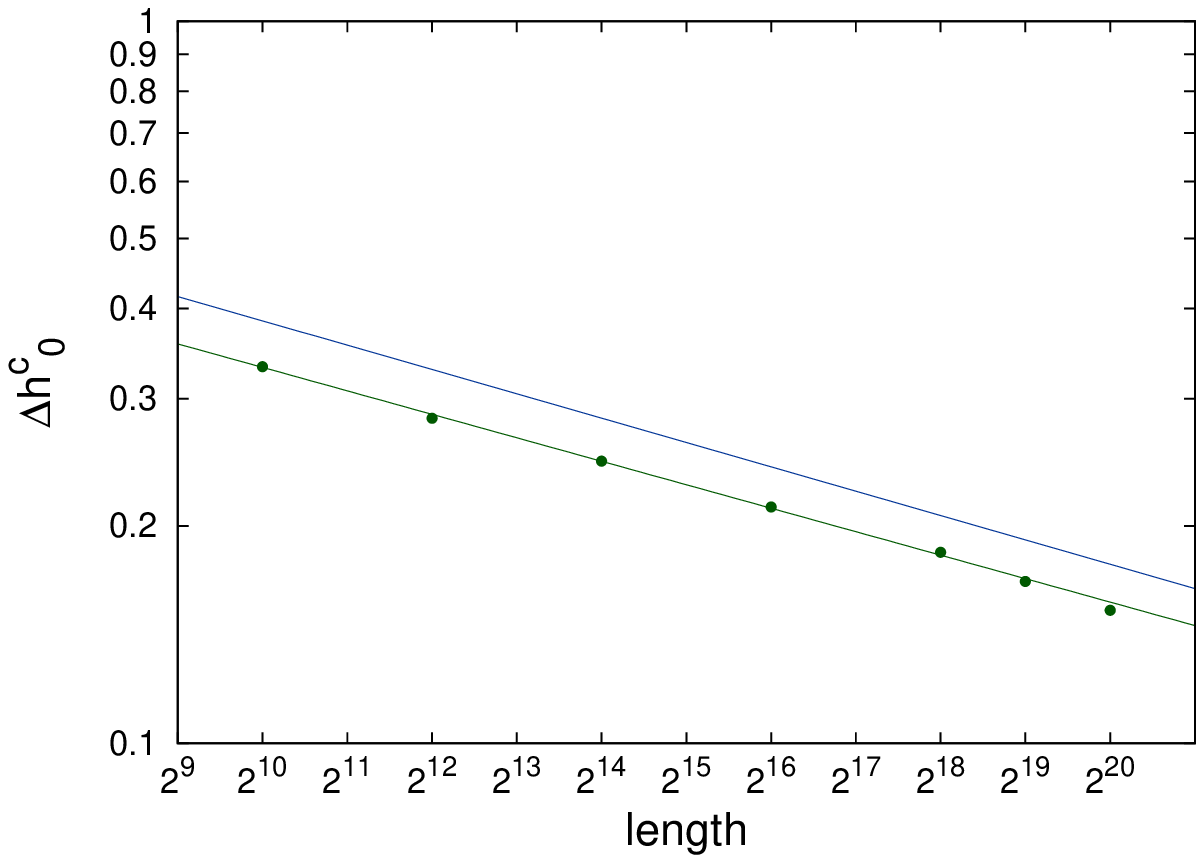}
}
\subfloat[][]{
\includegraphics[width=8truecm]{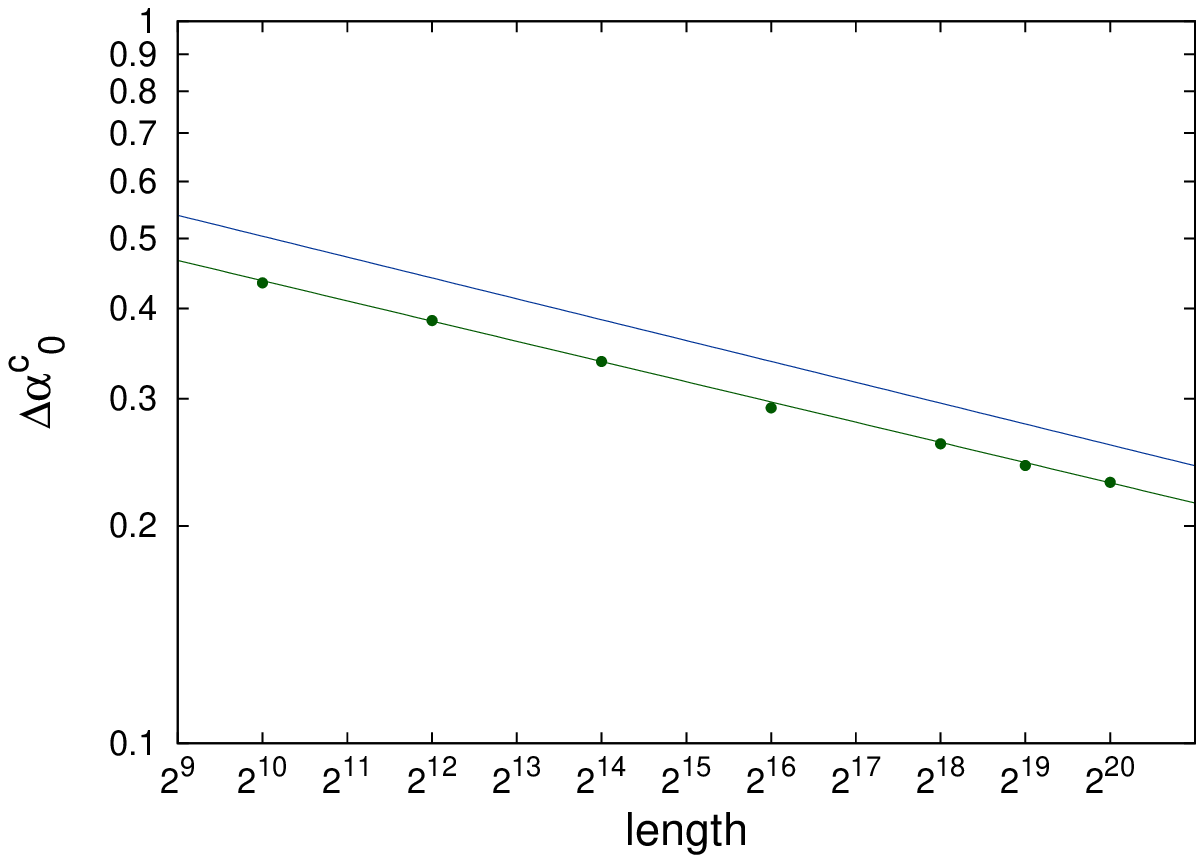}
}

\subfloat[][]{
\includegraphics[width=8truecm]{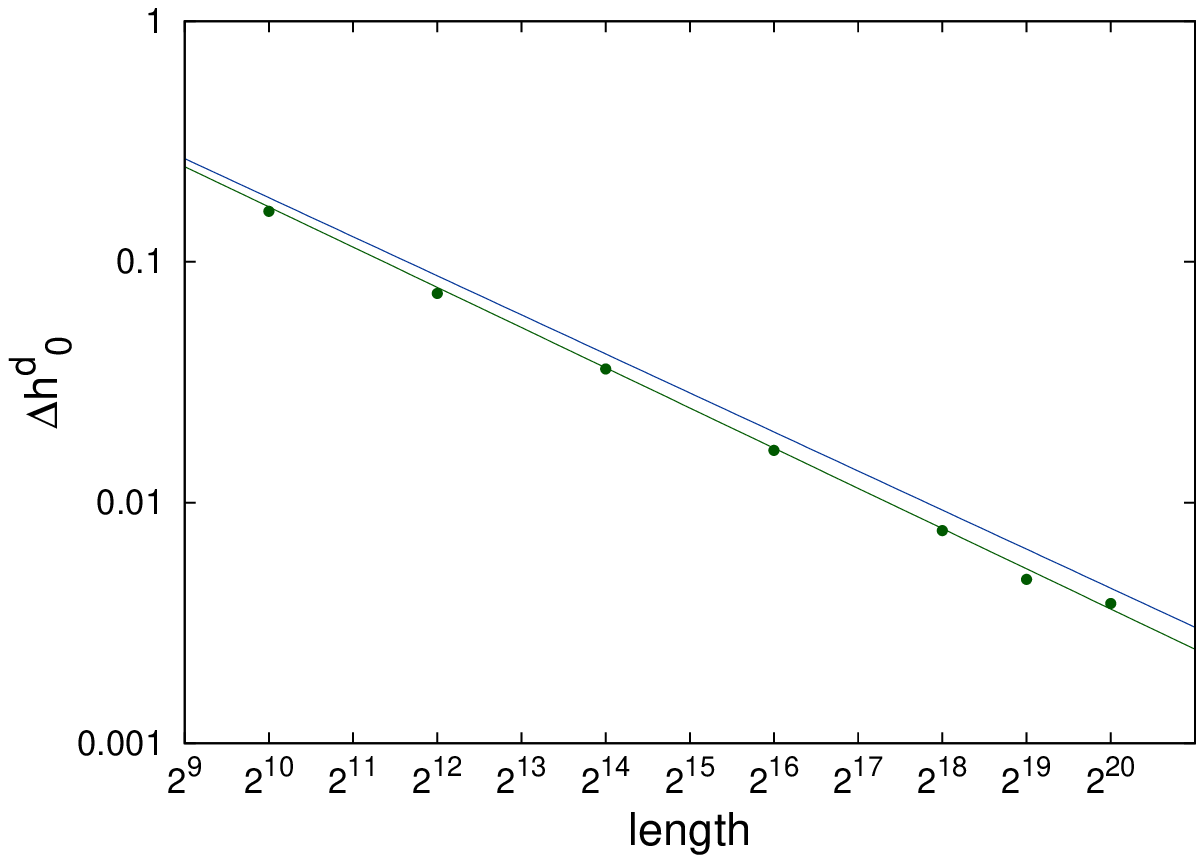}
}
\subfloat[][]{
\includegraphics[width=8truecm]{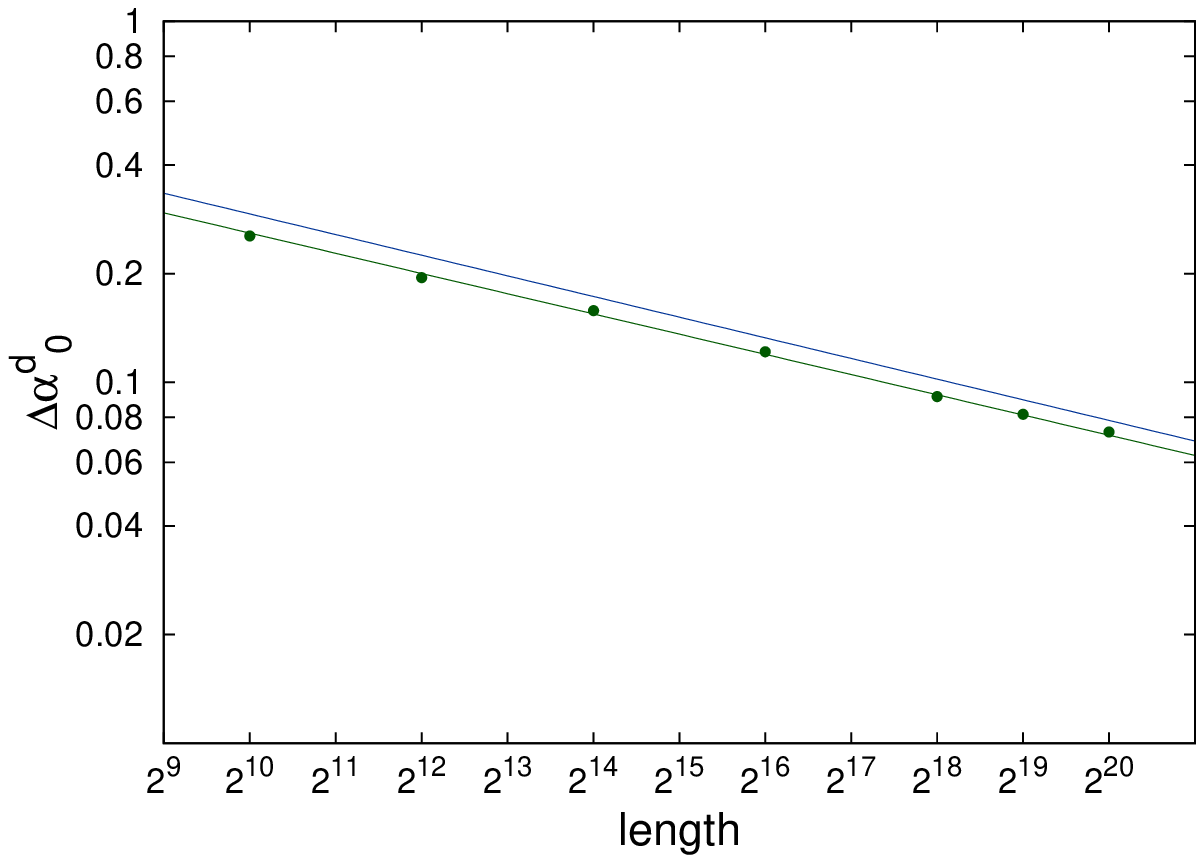}
}
\caption{Spread $\Delta h_0$ of generalized Hurst exponent (a)(c) and $\Delta\alpha_0$ of H\"older parameter (b)(d) for integrated (a)(b) differentiated (c)(d) fully persistent time series ($\gamma=0$) versus the length of data. The green line presents the power-law fit in log scale and the blue line corresponds to $95\%$ confidence level resulting from statistics. Fitted parameters are collected in Tables 3 -- 6.}
\label{}
\end{figure}
\clearpage

\begin{figure}[p]
\subfloat[][]{
\includegraphics[width=8truecm]{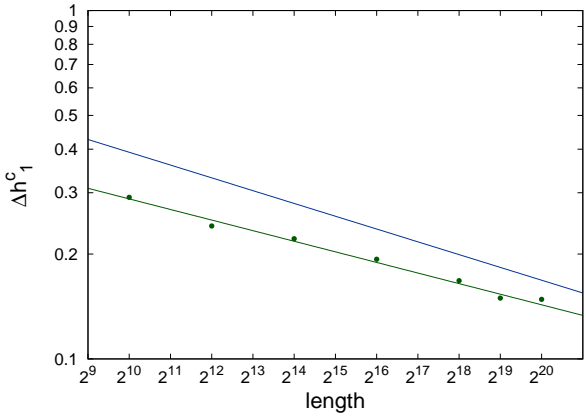}
}
\subfloat[][]{
\includegraphics[width=8truecm]{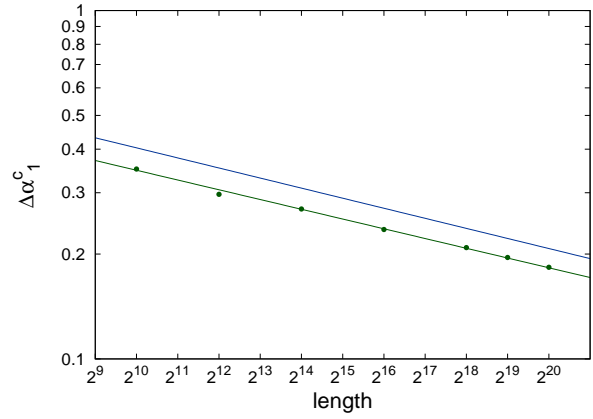}
}

\subfloat[][]{
\includegraphics[width=8truecm]{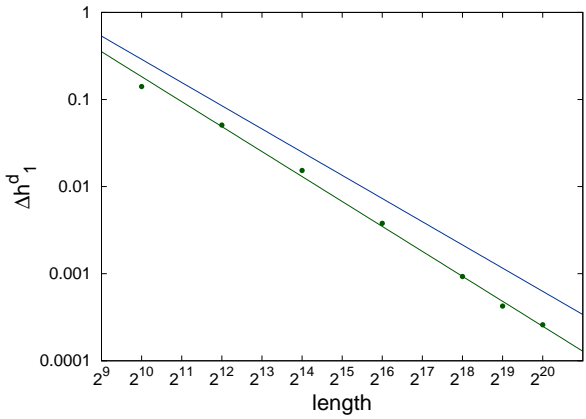}
}
\subfloat[][]{
\includegraphics[width=8truecm]{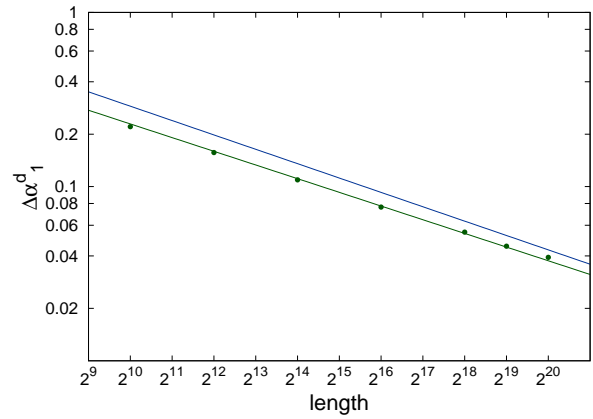}
}
\caption{Same as in Fig. 10, only for uncorrelated data ($\gamma=1$).}
\label{}
\end{figure}
\clearpage

\begin{figure}[p]
\subfloat[][]{
\includegraphics[width=8truecm]{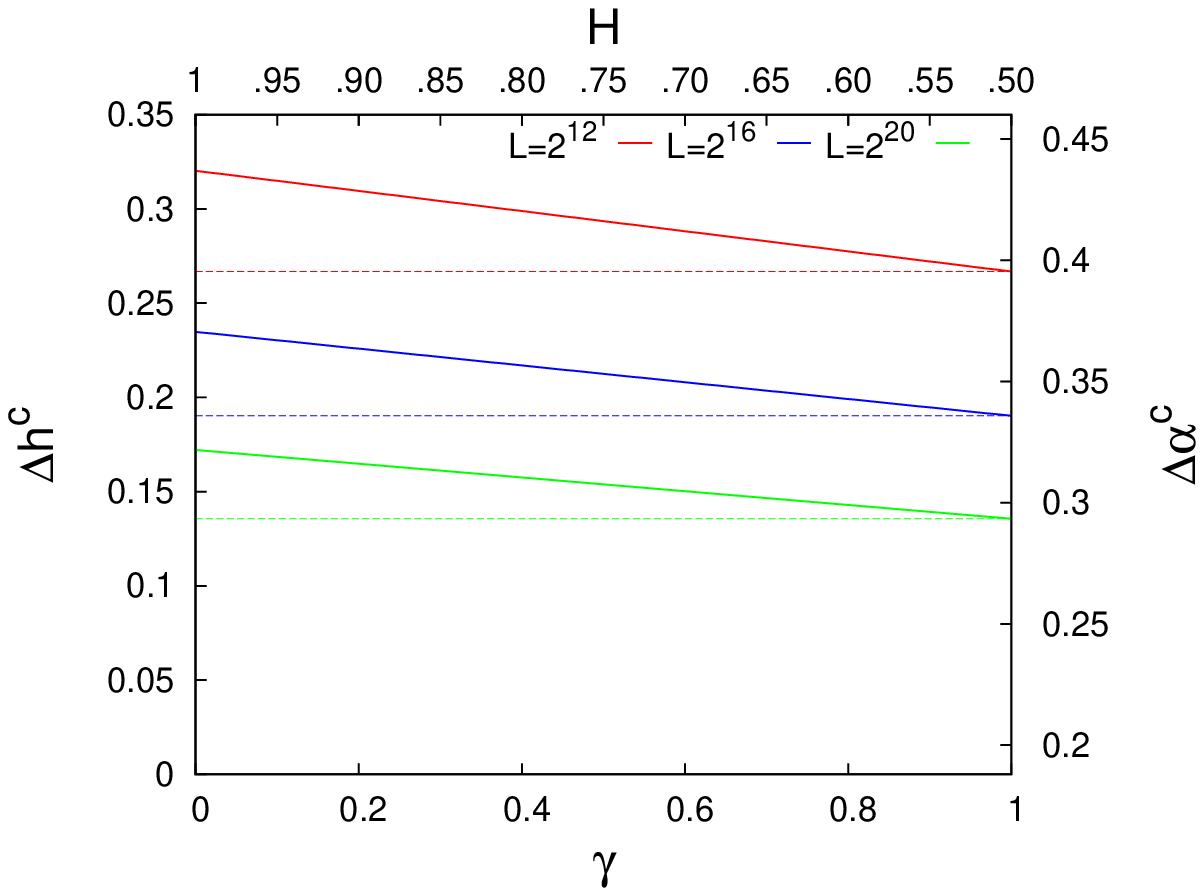}
}
\subfloat[][]{
\includegraphics[width=8truecm]{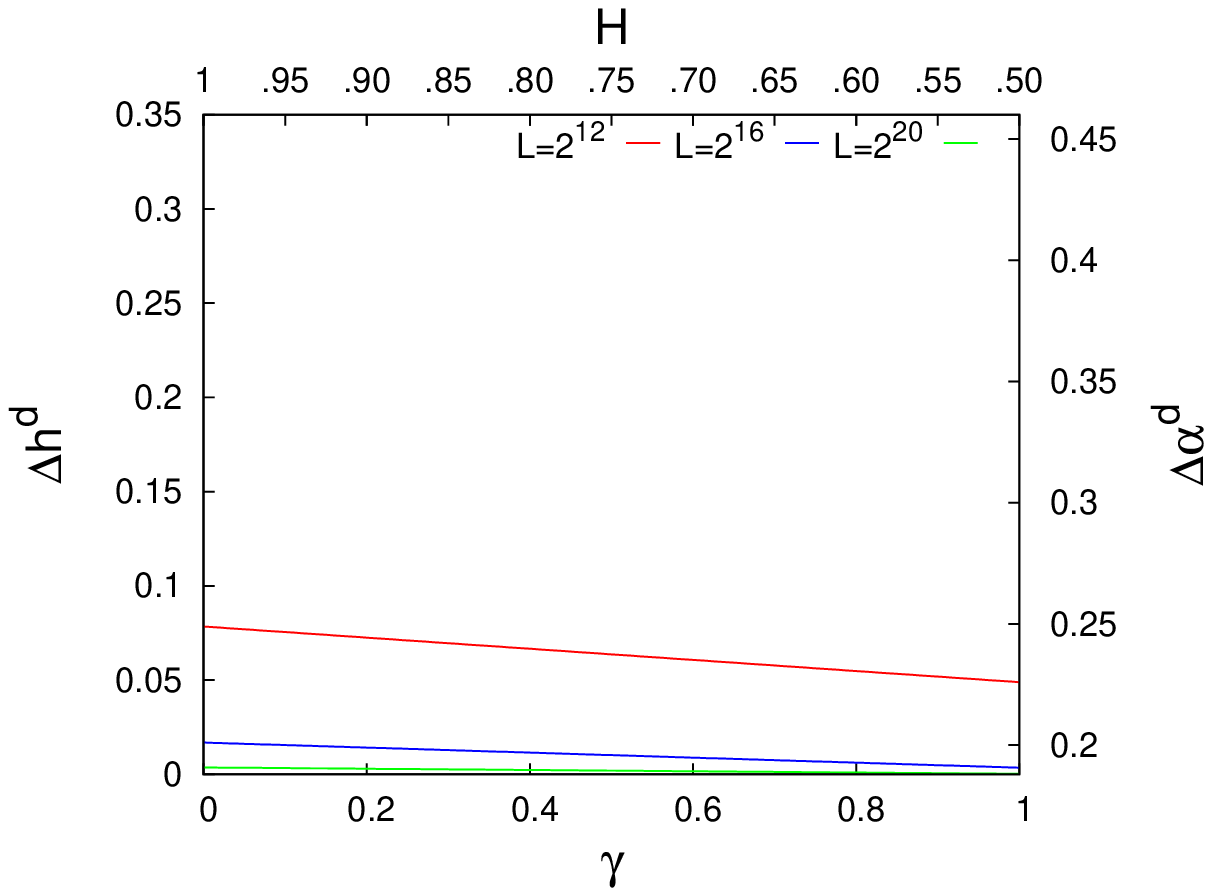}
}
\caption{Examples of $\Delta h$ and $\Delta \alpha$ FSE multifractal thresholds for integrated (a) and differentiated (b) time series. The notation is the same as in Fig.8.}
\label{h_sh-q}
\end{figure}
\clearpage

\begin{figure}[p]
\includegraphics[width=16truecm]{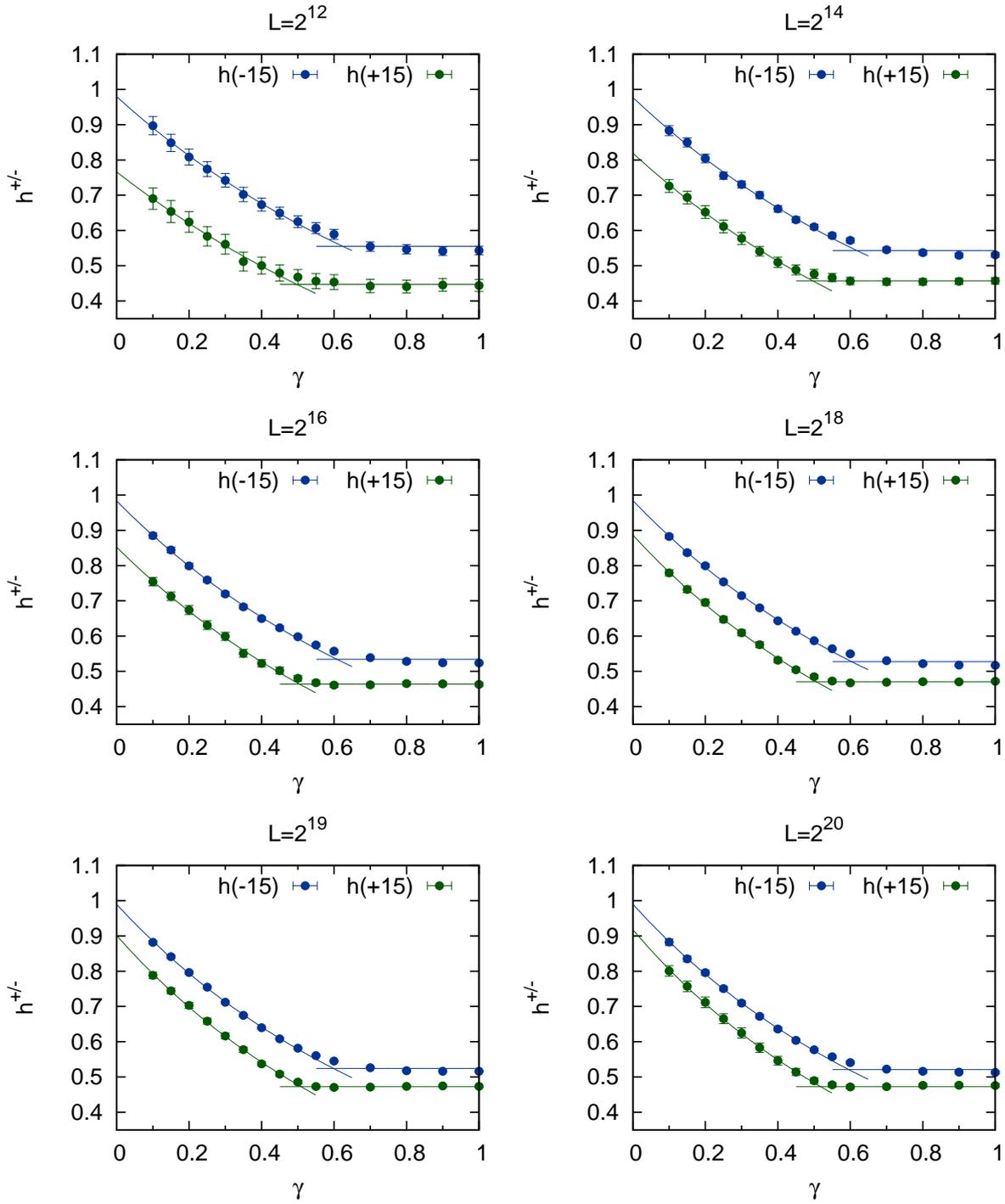}
\caption{Results for multifractal profile of data after $\Delta x_i \to |\Delta x_i|$ transformation. Plots of $h^{\pm}$ edge values of multifractal profiles are shown for transformed persistent time series of various lengths. An statistical ensemble of $5\cdot 10^2$ time series generated for each pair of length $L$ and persistency level $\gamma$ was taken in calculations and the resulting statistical uncertainty is marked for each $h^{\pm}$ value.}
\end{figure}
\clearpage
\begin{figure}[p]
\includegraphics[width=16truecm]{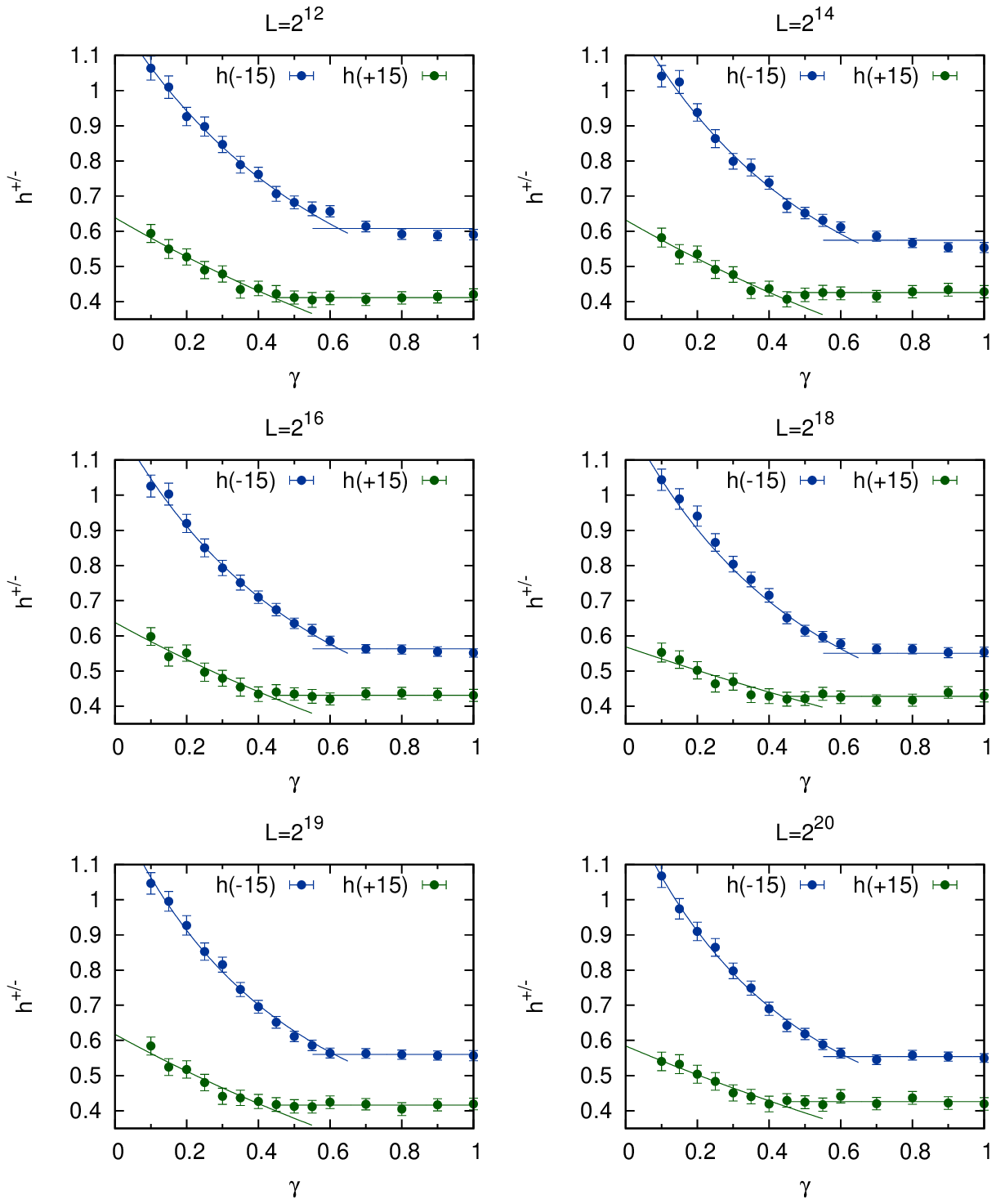}
\caption{Same as in Fig.13 but for $\Delta x_i \to (\Delta x_i)^2$ transformation.}
\end{figure}
\clearpage

\begin{figure}[p]

\subfloat[][$\Delta x_i\to|\Delta x_i|$]{
\includegraphics[width=8truecm]{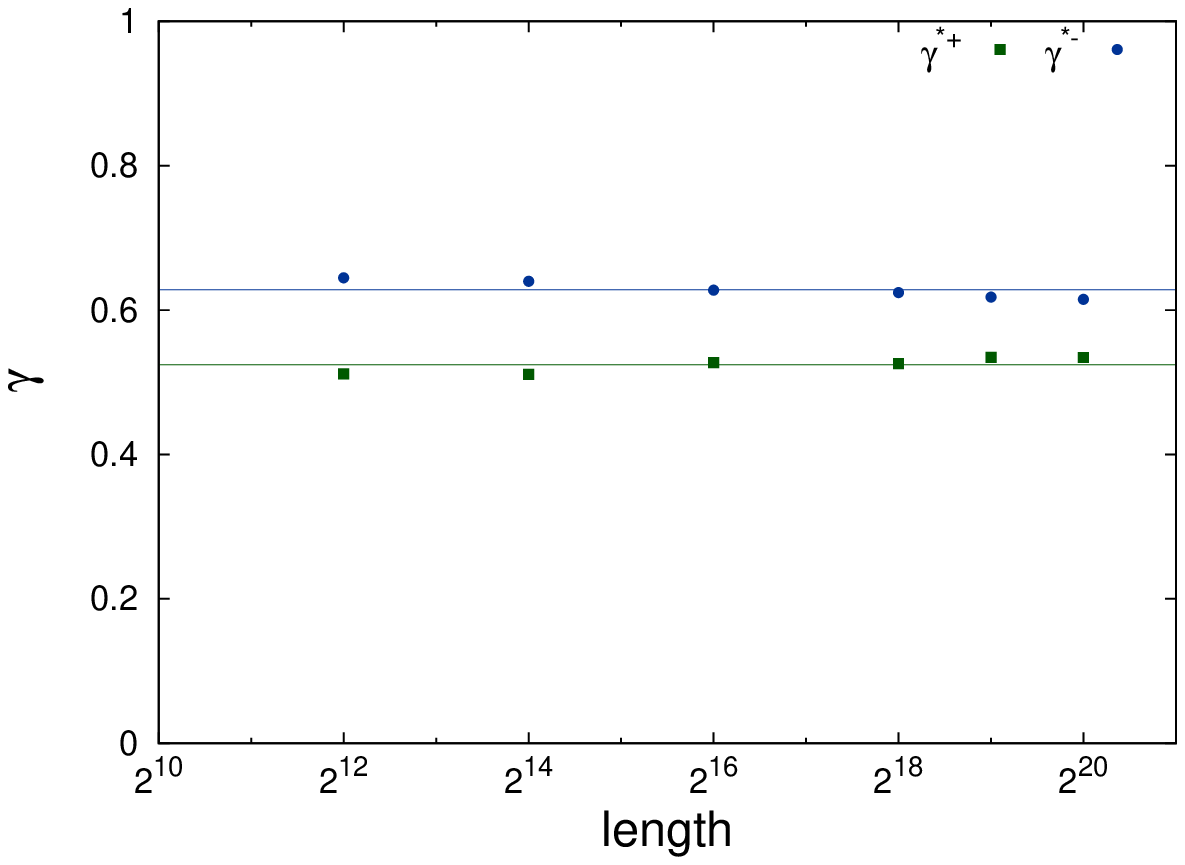}
}
\subfloat[][$\Delta x_i\to(\Delta x_i)^2$]{
\includegraphics[width=8truecm]{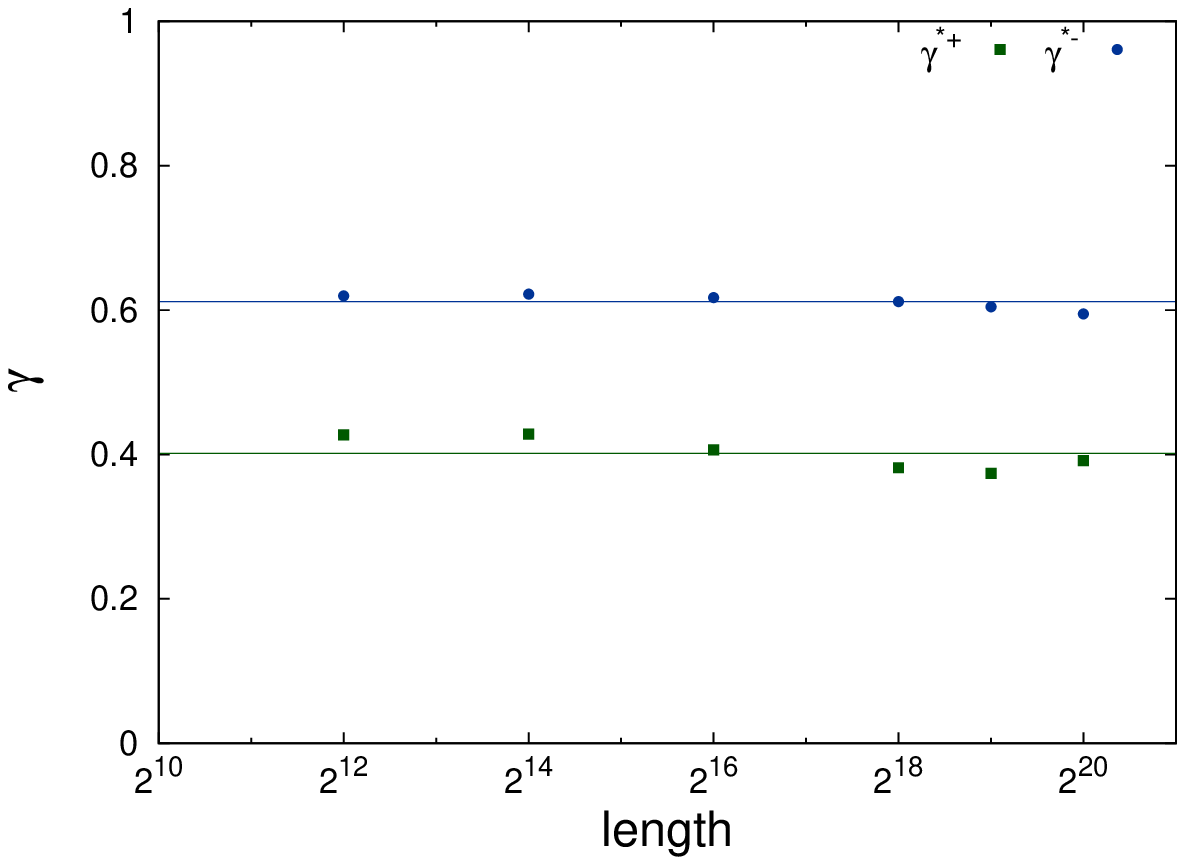}
}
\caption{Values of crossover points $\gamma^{*\pm}$ between regions of linear and nonlinear dependences of $h^{\pm}$ on autocorrelation scaling exponent $\gamma$. Parts (a) and (b) show this dependence against the length of data for $\Delta x_i \to |\Delta x_i|$ and $\Delta x_i \to (\Delta x_i)^2$ transformations respectively.}
\end{figure}
\clearpage

\begin{figure}[p]
\subfloat[][]{
\includegraphics[width=16truecm]{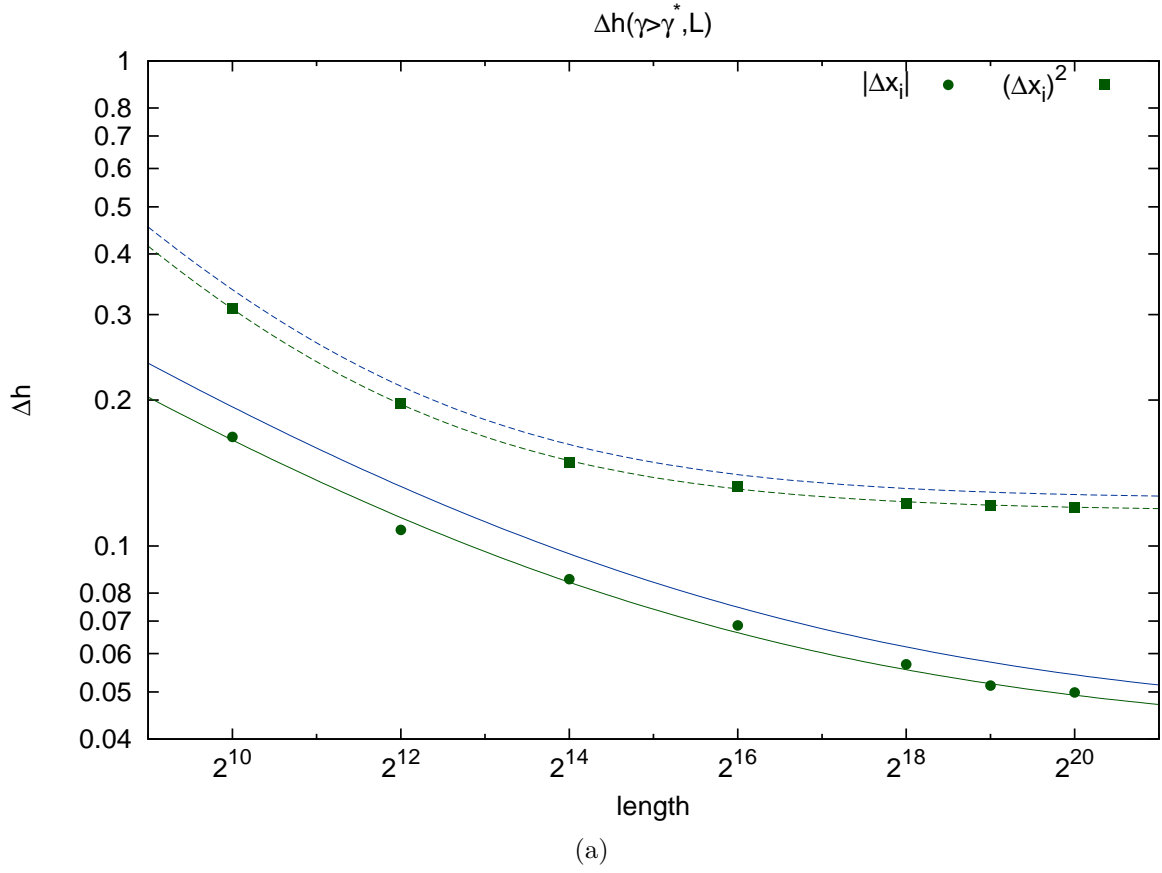}
}

\subfloat[][$\Delta x_i\to|\Delta x_i|$]{
\includegraphics[width=8truecm]{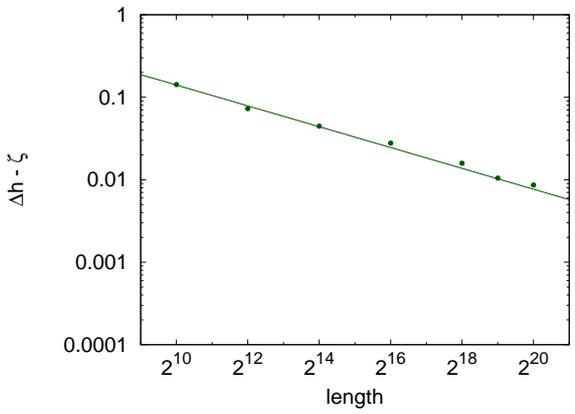}
}
\subfloat[][$\Delta x_i\to(\Delta x_i)^2$]{

\includegraphics[width=8truecm]{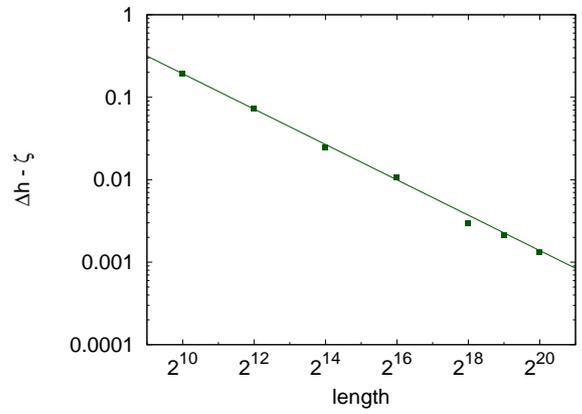}
}
\caption{Spread of multifractal profile $\Delta h$ presented as a function of data series length for region $\gamma>\gamma^*$. Plot (a) presents a fit of Eq.(27) while bottom plots (b) and (c) confirm in log-log scale the power law nature of $\Delta h-\zeta$ dependence on $L$ for both considered transformations. 
}
\end{figure}
\clearpage

\begin{figure}[p]
$$\Delta x_i\to|\Delta x_i|$$
\subfloat[][]{
\includegraphics[width=8truecm]{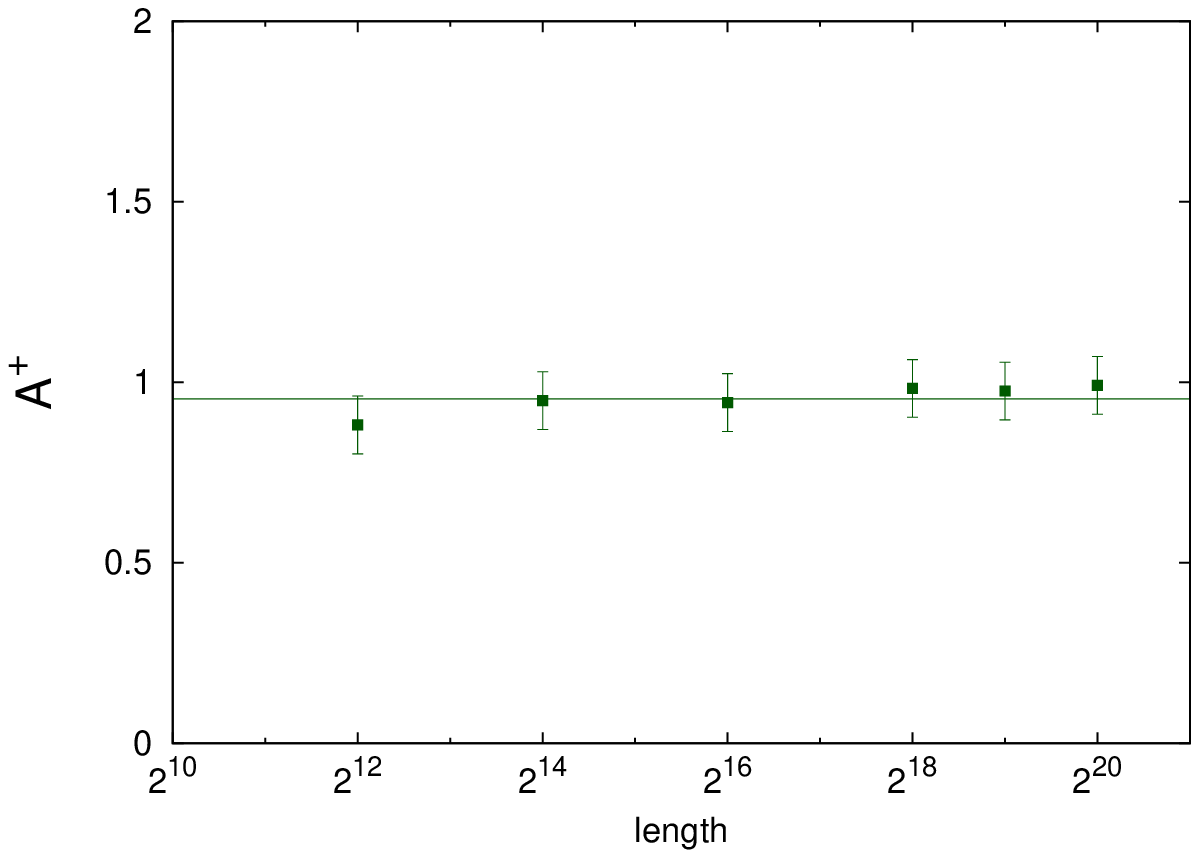}
}
\subfloat[][]{
\includegraphics[width=8truecm]{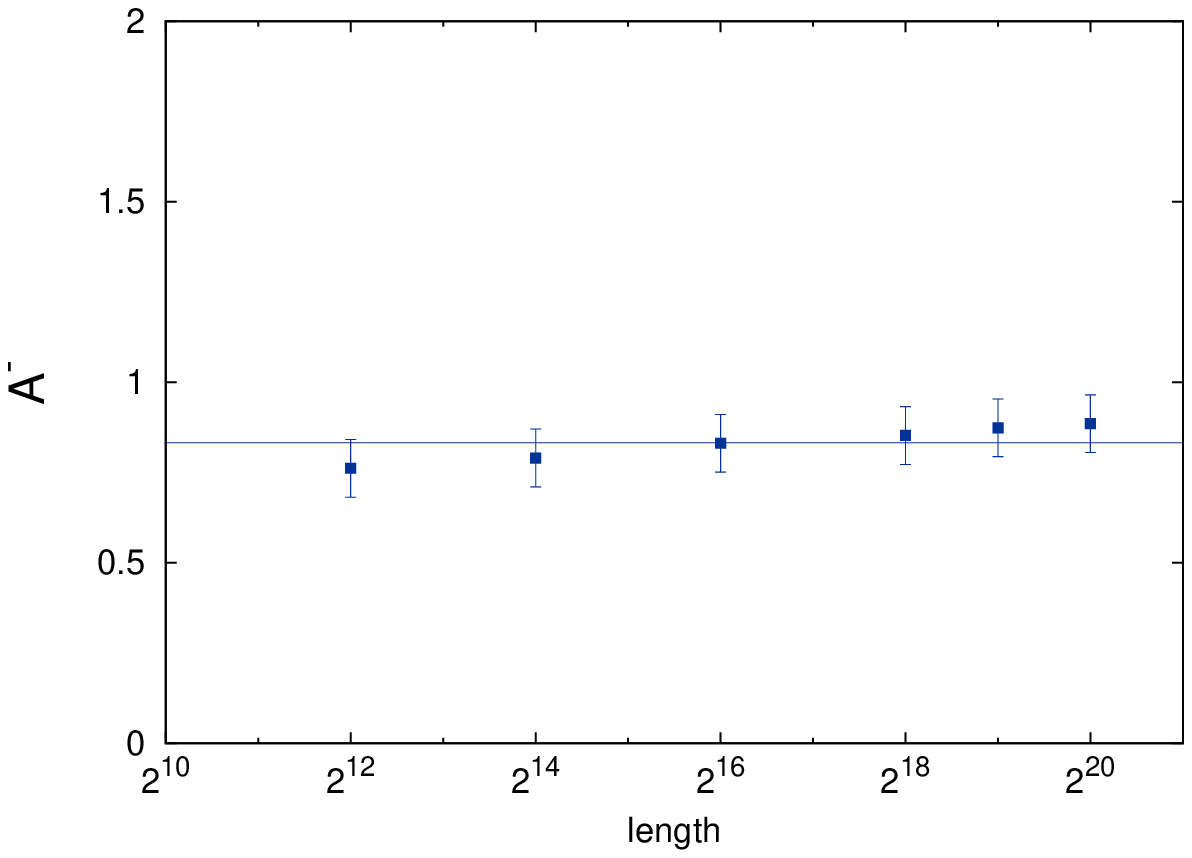}
}
\caption{Results for $A^{\pm}$ values of $q$-exponential fitting to the edges of multifractal profile of data after $\Delta x_i \to |\Delta x_i|$ transformation in the region $\gamma<\gamma^*$ (see Eq.(28)). The plot (a) shows dependence of $A^{+}$ on $L$ while the plot (b) reveals this relation for $A^-$ values. The very weak dependence on $L$ in log--linear scale proves that both parameters can be assumed constant. The error bars indicate fit uncertainties.
}
\end{figure}
\clearpage

\begin{figure}[p]
$$\Delta x_i\to(\Delta x_i)^2$$
\subfloat[][]{
\includegraphics[width=8truecm]{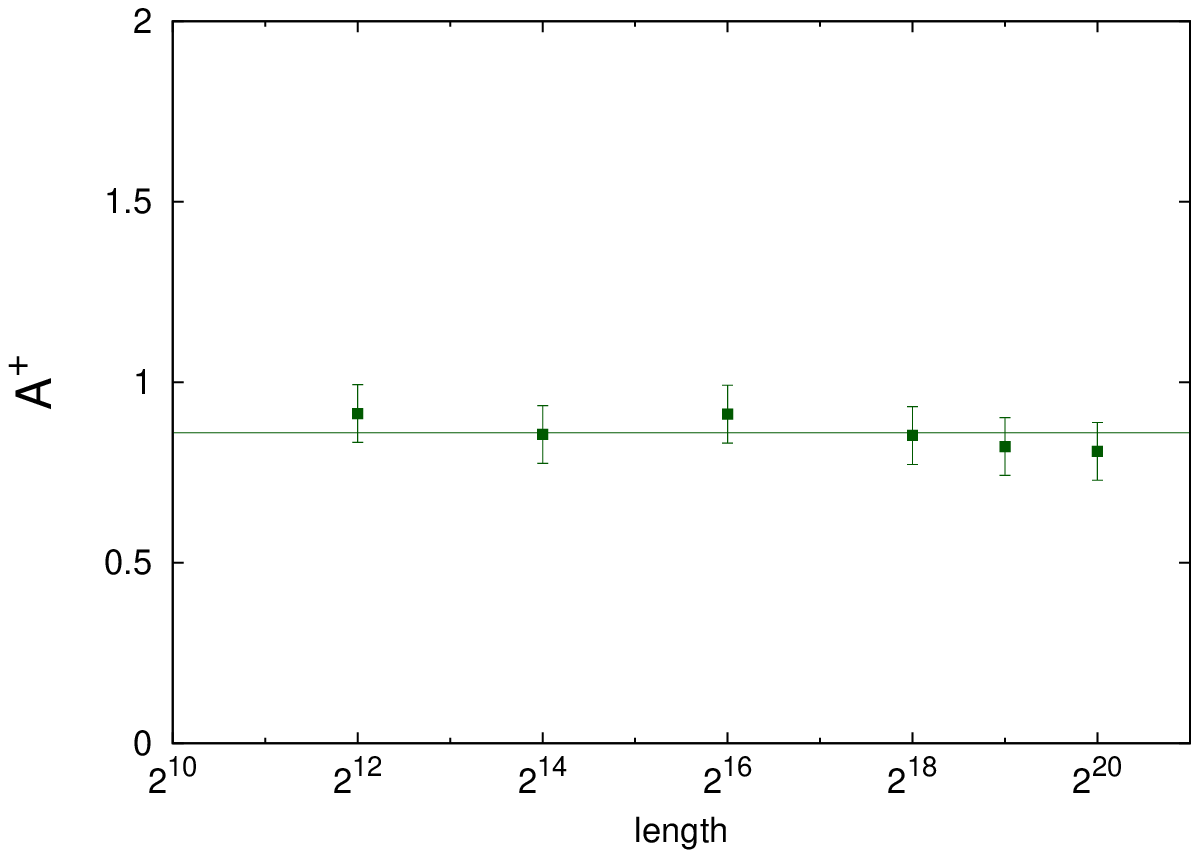}
}
\subfloat[][]{
\includegraphics[width=8truecm]{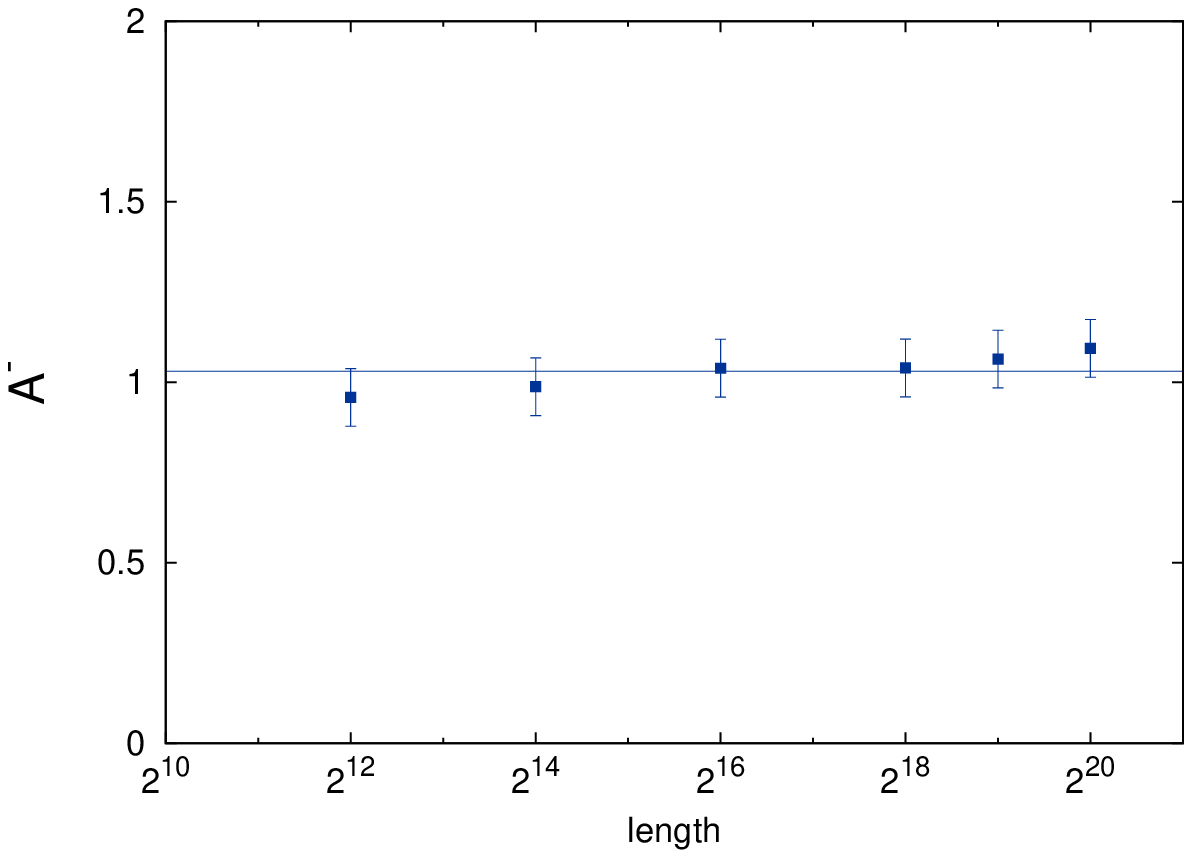}
}
\caption{Same as in Fig.17 but for $\Delta x_i \to (\Delta x_i)^2$ transformation.}
\end{figure}
\clearpage

\begin{figure}[p]
\subfloat[][$\Delta x_i\to|\Delta x_i|$]{
\includegraphics[width=8truecm]{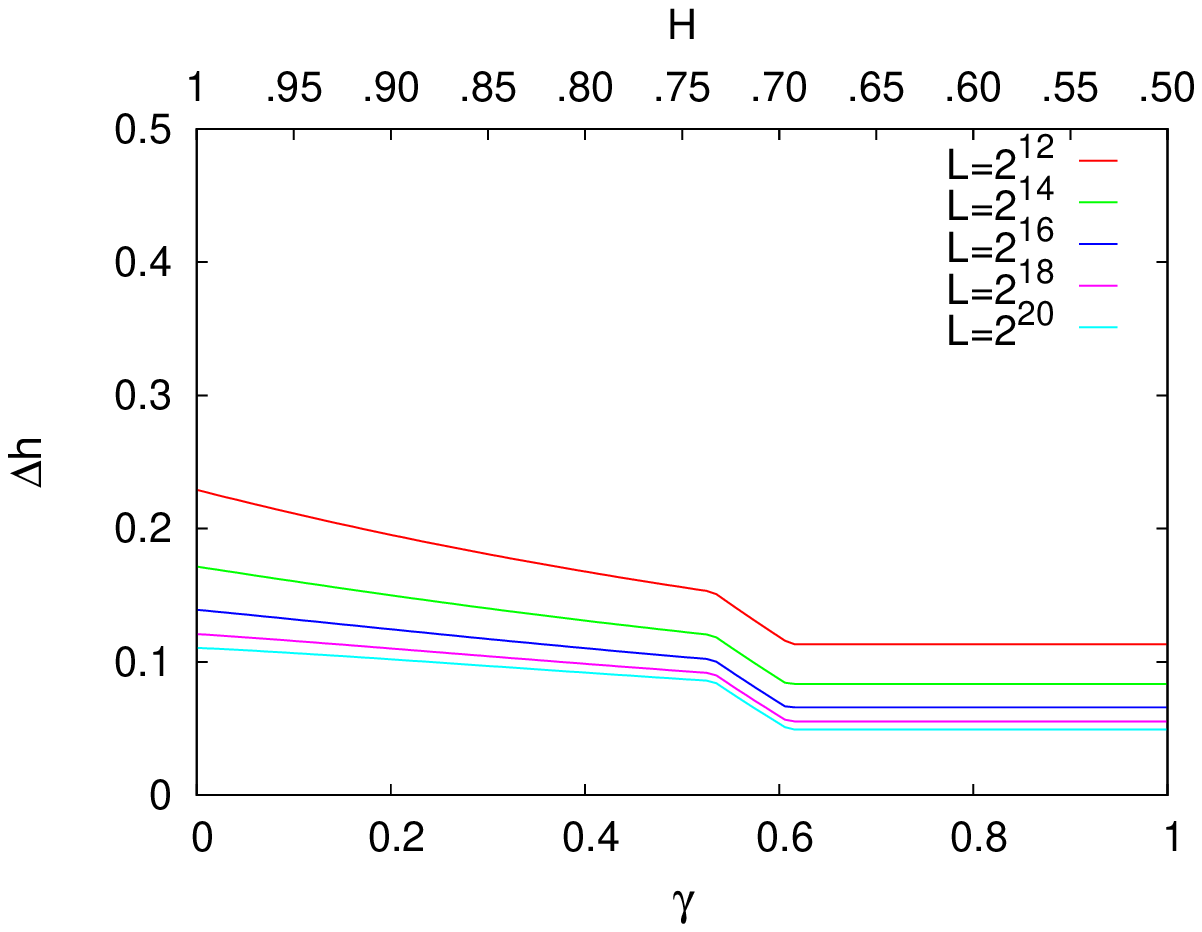}
}
\subfloat[][$\Delta x_i\to(\Delta x_i)^2$]{
\includegraphics[width=8truecm]{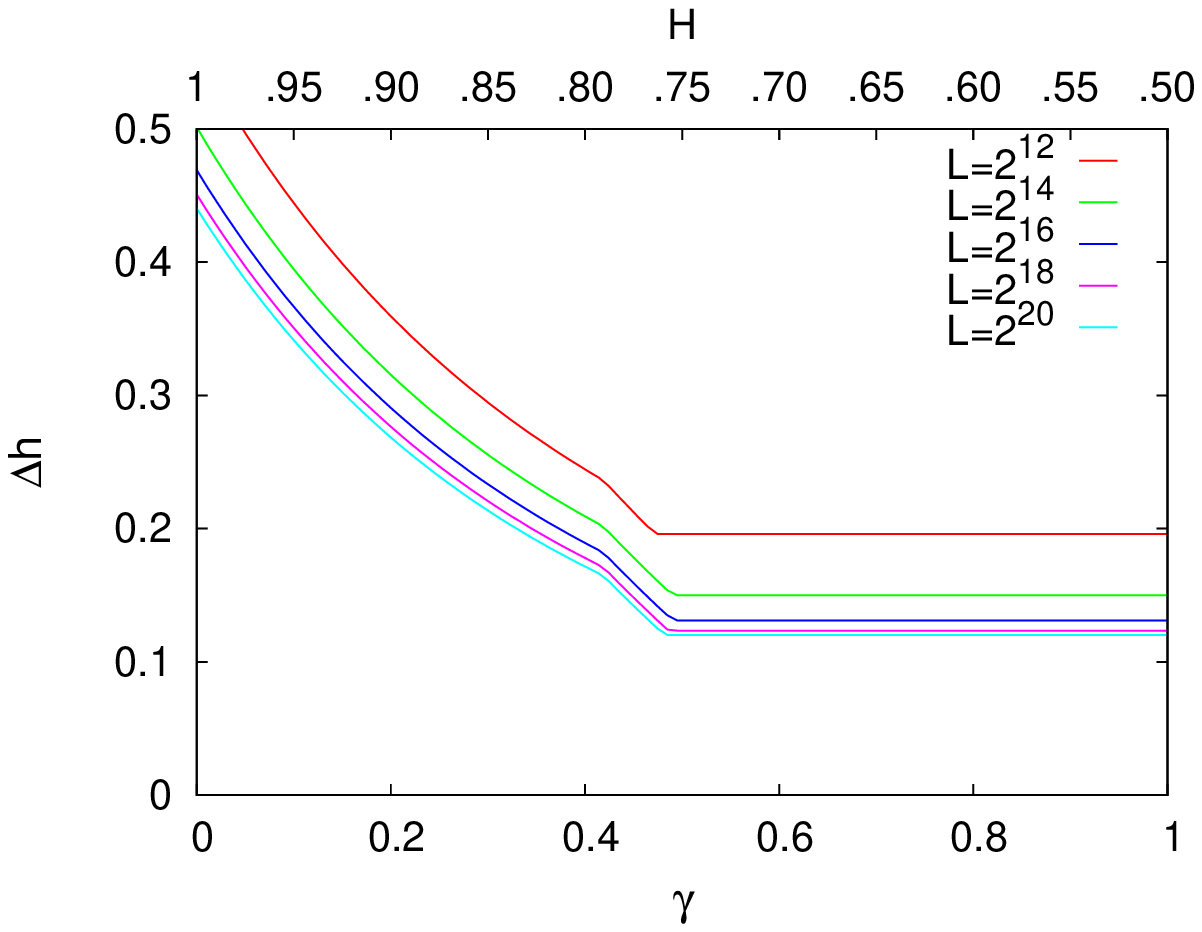}
}
\caption{Examples of $\Delta h$ multifractal spread originated from (a) $\Delta x_i \to |\Delta x_i|$ and (b)  $\Delta x_i \to (\Delta x_i)^2$ transformations for variety of data lenghts $L= 2^{12}, 2^{14}, 2^{16}, 2^{18}, 2^{20}$ drawn as a function of persistency level $(\gamma)$ in primary data.  The respective values of Hurst exponent are marked on top axis. Two different behaviors in partly overlapping regions  are visible -- one with nonlinear descent for $\gamma<\gamma^{*+}$ (left) and the second with $\Delta h (\gamma) = const$ for $\gamma>\gamma^{*-}$ (right). The transition area for $\gamma^{*+}<\gamma<\gamma^{*-}$ is also noticeable.}
\end{figure}
\clearpage

\end{document}